\documentclass[transmag]{IEEEtran}
\DeclareUnicodeCharacter{202A}{\,}
\DeclareUnicodeCharacter{202C}{\,}

\usepackage{subcaption}
\usepackage{amsmath,amsthm,amssymb,amsfonts}
\usepackage[hidelinks]{hyperref}
\usepackage{bm}

\theoremstyle{remark}

\newcommand{\bmA}{\mathbf A}

\newcommand{\bmI}{\mathbf I}
\newcommand{\bmT}{\mathbf T}
\newcommand{\bmQ}{\mathbf Q}
\newcommand{\bmB}{\mathbf B}

\newcommand{\bmE}{\mathbf E}
\newcommand{\bmL}{\mathbf L}

\newcommand{\bmH}{\mathbf H}
\newcommand{\bmF}{\mathbf F}
\newcommand{\bmS}{\mathbf S}
\newcommand{\bms}{\mathbf s}
\newcommand{\bmV}{\mathbf V}
\newcommand{\bmU}{\mathbf U}

\newcommand{\bmK}{\mathbf K}
\newcommand{\bmy}{\mathbf y}
\newcommand{\bmZ}{\mathbf Z}
\newcommand{\bmJ}{\mathbf J}
\newcommand{\bmR}{\mathbf R}
\newcommand{\bmX}{\mathbf X}
\newcommand{\bmW}{\mathbf W}

\newcommand{\bmq}{\mathbf q}

\newcommand{\bmx}{\mathbf x}
\newcommand{\bmn}{\mathbf n}

\usepackage{xspace}
\usepackage{amssymb}
\usepackage{graphicx}
\usepackage{amsmath}
\usepackage{mathtools}
\theoremstyle{plain}
\usepackage{graphicx} 
\usepackage{amsmath, amsfonts, amssymb, amsbsy,nccmath} 
\usepackage{algorithm} 

\usepackage{enumerate} 
\usepackage{algorithmic} 
\usepackage{lipsum} 
\usepackage[sort,compress]{cite} 
\usepackage{epsfig} 
\usepackage{epstopdf}
\usepackage{mathtools}
\usepackage{dsfont} 
\usepackage[inline]{enumitem}
\usepackage{color,soul}

\usepackage{stfloats}  
\usepackage{tabularx}
\usepackage{xcolor}
\usepackage{lipsum}
\usepackage{mwe}
\usepackage{setspace}
\usepackage{longtable}
\usepackage{subcaption}
\usepackage{caption}
\usepackage[normalem]{ulem}
\usepackage[user]{zref}
\newcommand{\sot}[1]{} %\textcolor{red}{\sout{#1}}}

\usepackage{xpatch}

\newcounter{revc}
\makeatletter \zref@newprop{revcontent}{} \zref@addprop{main}{revcontent}
\zref@newprop{revsec}{} \zref@addprop{main}{revsec}
\zref@newprop{revpage}{} \zref@addprop{main}{revpage}

\newcommand{\revi}[2]{
\zref@setcurrent{revsec}{\thesection}%
\zref@setcurrent{revpage}{\thepage}%
\zref@setcurrent{revcontent}{#2}%
\refstepcounter{revc}%
\label{#1}%
\zlabel{#1}%
\textcolor{blue}{#2}%
}

\newcommand{\revinu}[2]{%
\zref@setcurrent{revsec}{\thesection}%
\zref@setcurrent{revcontent}{#2}%
\refstepcounter{revc}%
\zlabel{#1}%
\label{#1}
#2 }

\newcommand{\revr}[2]{%
\zref@setcurrent{revsec}{\thesection}%
\zref@setcurrent{revcontent}{#2}%
\refstepcounter{revc}%
\zlabel{#1}%
\label{#1} \sot{#2}} \makeatother

\expandafter\def\expandafter\quote\expandafter{\quote\onehalfspacing\fontsize{12}{14}\selectfont}

\usepackage[framemethod=tikz]{mdframed}
\usepackage{lipsum}

\definecolor{mycolor}{rgb}{0.122, 0.435, 0.698}

\newmdenv[innerlinewidth=0.5pt, roundcorner=4pt,linecolor=mycolor,innerleftmargin=6pt,
innerrightmargin=6pt,innertopmargin=6pt,innerbottommargin=6pt]{mybox}

\def\BibTeX{{\rm B\kern-.05em{\sc i\kern-.025em b}\kern-.08em T\kern-.1667em\lower.7ex\hbox{E}\kern-.125emX}}
\begin{document}
\sloppy
\title{IRS Assisted MIMO Full Duplex: Rate Analysis and Beamforming Under Imperfect CSI}

%\author{Chandan Kumar Sheemar, \IEEEmembership{Student Member, IEEE}, Christo Kurisummoottil Thomas, \IEEEmembership{Member, IEEE}, \\ and Dirk Slock, \IEEEmembership{Fellow, IEEE}

%\thanks{Chandan Kumar Sheemar and Dirk Slock are with the Communication Systems Department at EURECOM, Sophia Antipolis, 06410, France (emails:sheemar@eurecom.fr,slock@eurecom.fr);  }
%\thanks{Christo Kurisummoottil Thomas is with Qualcomm Finland RFFE Oy, Keilaranta 8, 02150 Espoo
%(e-mail: ckurisum@qti.qualcomm.com).}
%\thanks{This work was financially supported by EURECOM.}
%}

\author{Chandan~Kumar~Sheemar, \IEEEmembership{Member~IEEE,} ‪Sourabh Solanki‬, \IEEEmembership{Member~IEEE,}\\
   \;Jorge Querol,~\IEEEmembership{Member~IEEE,} 
   Sumit Kumar,~\IEEEmembership{Member~IEEE,}
Symeon Chatzinotas,~\IEEEmembership{Fellow,~IEEE}  % <-this % stops a space
  %    \thanks{An initial version of this work is part of the PhD thesis  \cite{sheemar2022hybrid_thesis}.}% <-this % stops a space
\thanks{The authors are with the Interdisciplinary Centre for Security Reliability and Trust (SnT), University of Luxembourg (emails:\{chandankumar.sheemar, sourabh.solanki, jorge.querol, sumit.kumar, symeon.chatzinotas\}uni.lu).}% <-this % stops a space
\thanks{A preliminary version of this work will appear in IEEE GLOBECOM 2023.}
}

\IEEEtitleabstractindextext{\begin{abstract}
Intelligent reflecting surfaces (IRS) have emerged as a promising technology to enhance the performance of wireless communication
systems. By actively manipulating the wireless propagation environment, IRS enables efficient signal transmission and reception. In
recent years, the integration of IRS with full-duplex (FD) communication has garnered significant attention due to its potential to
further improve spectral and energy efficiencies. IRS-assisted FD systems combine the benefits of both IRS and FD technologies,
providing a powerful solution for the next generation of cellular systems. In this manuscript, we present a novel approach to jointly
optimize active and passive beamforming in a multiple-input-multiple-output (MIMO) FD system assisted by an IRS for weighted
sum rate (WSR) maximization. Given the inherent difficulty in obtaining perfect channel state information (CSI) in practical
scenarios, we consider imperfect CSI and propose a statistically robust beamforming strategy to maximize the ergodic WSR. Additionally, we analyze the achievable WSR for an IRS-assisted MIMO FD system under imperfect CSI by deriving
both the lower and upper bounds. To tackle the problem of ergodic WSR maximization, we employ the concept of expected weighted
minimum mean squared error (EWMMSE), which exploits the information of the expected error covariance matrices and ensures convergence to a local optimum. We evaluate the effectiveness of our 
proposed design through extensive simulations. The results demonstrate that our robust approach yields significant performance
improvements compared to the simplistic beamforming approach that disregards CSI errors, while also outperforming the robust
half-duplex (HD) system considerably.
\end{abstract}

}
 
\maketitle
\section{Introduction}  
\IEEEPARstart{F}{ull Duplex} (FD) is a promising wireless transmission technology offering simultaneous transmission and reception in the same frequency band, which theoretically doubles the spectral efficiency \cite{sabharwal2014band,krikidis2012full}.
It is crucial for sustaining the exponentially ever-increasing data rate demands as it can offer flexible utilization of the limited wireless spectrum \cite{cirik2014achievable,sheemar2021per}. Beyond spectral efficiency, FD can be beneficial to improve security, enable advanced joint communication and sensing, and reduce end-to-end delays \cite{zhu2014joint,islam2022integrated,barneto2021full}. However, a major hurdle in achieving optimal FD operation is the presence of self-interference (SI) which can reach up to $90-110$ dB higher than the desired received signal \cite{duarte2012experiment, sheemar2022practical}. Overcoming this challenge is crucial for realizing the full potential of FD systems.

%However, FD systems suffer from Self-Interference (SI), which could be $90-110~$dB higher than the received signal of interest \cite{duarte2012experiment,sheemar2022practical}, and it is a major challenge for achieving an ideal FD operation.

Sophisticated self-interference cancellation (SIC) techniques hold paramount significance in effectively reducing the power of SI to levels approaching the noise floor thereby enabling precise reception of the desired signal. These techniques can be categorized into two primary classifications: passive and active SIC \cite{duarte2012experiment,aquilina2017weighted}. Passive SIC involves attenuating SI by optimizing the path loss between the transmit and receive antennas of the FD node. Through strategic manipulation of signal propagation, passive SIC aims to effectively diminish the power of SI. Active SIC techniques can be further divided into two distinct domains: analog and digital techniques \cite{alexandropoulos2020full, riihonen2012analog, mohammadi2016throughput}. Analog SIC primarily focuses on mitigating SI by replicating the transmitted signal with an inverted sign prior to reaching the analog-to-digital converters (ADCs). This approach ensures that an adequate dynamic range is preserved in the ADCs for the desired signal. Subsequently, the residual SI that remains after the analog-to-digital conversion stage is addressed using digital SIC techniques in the baseband \cite{ahmed2015all}. The combined utilization of active and passive SIC techniques allows for the substantial reduction of SI near the noise floor, resulting in a twofold increase in spectral efficiency.

Besides FD technology, a promising and emerging innovation called intelligent reflecting surface (IRS) is garnering considerable attention \cite{hu2018beyond,chen2016review}. These surfaces provide a dynamic and adaptable wireless environment, offering full control over the manipulation of the impinging electromagnetic field to fulfill specific requirements \cite{Solanki2022OJCOMS}. Constructed from multiple cost-effective passive meta-elements, IRSs consist of planar surfaces capable of modifying the reflected signal without the need for a dedicated radio frequency (RF) chain. As a result, IRSs can be deployed with significantly lower energy costs compared to traditional active nodes. Through intelligent reconfiguration, an IRS can fulfill various functions, including enhancing signal reception quality by bypassing obstacles, compensating for signal fading, introducing supplementary signal paths, improving channel statistics, and fortifying wireless network security \cite{wu2021intelligent}.

%In addition to FD technology, there is another emerging innovation called intelligent reflecting surfaces (IRSs) that is gaining traction \cite{hu2018beyond,chen2016review}. These surfaces offer a dynamic and adaptable wireless environment, granting complete control over the manipulation of the impinging electromagnetic field to meet specific requirements. IRSs consist of planar surfaces constructed from numerous cost-effective passive meta-elements capable of altering the reflected signal without the need for a radio frequency (RF) chain. Consequently, IRSs can be implemented with significantly lower energy costs compared to traditional active nodes. Leveraging smart reconfiguration, an IRS can perform various functions including enhancing signal reception quality by circumventing obstacles, compensating for fading, introducing supplementary signal paths, refining channel statistics, and bolstering wireless network security   \cite{wu2021intelligent}.}

The recent literature on the IRSs is available in \cite{basar2021present,jiang2023physics,jiang2023hybrid}
In \cite{basar2021present}, the recent developments in the different types of IRSs and promising candidates for future
research are discussed. In \cite{jiang2023physics}, the authors investigate the three-dimensional physics-based double IRSs for the unmanned aerial vehicle-to ground communication scenarios. A novel stochastic channel model
is proposed and the critical propagation properties of the proposed
channel model, such as the spatial cross-correlation functions, temporal auto-correlation functions, and frequency correlation functions, with respect to different RIS reflection phase configurations are investigated. In \cite{jiang2023hybrid}, a hybrid near-field and far-field stochastic channel model
for characterizing an IRS-assisted vehicle-to-vehicle
propagation environment is proposed. The potential of IRSs combined with  the upcoming sixth-generation (6G) \cite{saad2019vision} cellular networks can pave the way towards smartly controlled and energy-efficient wireless systems.

\subsection{State-of-the-Art on IRS Assisted FD systems}

The integration of the IRSs with FD (IRS-FD) systems, presents a highly promising solution for meeting the escalating traffic demands of the future. These systems offer the potential to achieve optimal utilization of available resources, encompassing both spectrum and power allocation \cite{liaskos2018new,foo2017liquid,sheemar2022hybrid}. By leveraging the intelligent capabilities of IRS and the simultaneous transmission and reception capabilities of FD, IRS-FD systems are poised to revolutionize wireless communication networks. This innovative approach has garnered significant attention from researchers and experts in the field, with studies highlighting the transformative potential of IRS-FD in addressing the evolving requirements of modern wireless communication systems \cite{sheemar2023full}.

A collection of recent studies exploring IRS-FD systems can be found in \cite{sharma2020intelligent,abdullah2020optimization,peng2021multiuser,cai2021intelligent,sheemar2022near,ge2021robust,liu2021deep,saeidi2021weighted}.
In \cite{sharma2020intelligent}, the authors investigated the performance of the IRS-FD system by analyzing the outage and error probabilities. Moreover, they also investigated the advantage of an IRS in partially reducing the SI effect. In \cite{abdullah2020optimization}, the authors presented a novel beamforming design for an FD relay assisted with one IRS to maximize the minimum achievable rate with the $\max-\min$ optimization. In \cite{peng2021multiuser}, the advantage of IRS in covering the dead zones in a multi-user FD system in investigated. The problem of minimization of the uplink (UL) and downlink (DL) power consumption under the minimum rate constraints for an IRS-FD system is analyzed in \cite{cai2021intelligent}.  In \cite{sheemar2022near}, beamforming for point-to-point IRS-FD is investigated.
In \cite{ge2021robust}, the authors studied the advantage of the IRS in improving the security of the FD systems and presented a novel beamforming design to maximize the worst-case achievable security rate.
The authors in \cite{liu2021deep} explored a mixed time-scale beamforming design for an IRS-assisted multi-user FD system. In addition, a deep neural network is also developed to reduce the computational burden of the proposed beamforming solution. Finally, in \cite{saeidi2021weighted}, the authors proposed a 
joint beamforming design for weighted sum rate (WSR) maximization in a single-cell multiple-input single-output (MISO) IRS-FD system.
 
%Despite the fruitful results on IRS-FD systems, the studies are limited to simple point-to-point MIMO FD systems or single antenna users case. Moreover, these studies consider perfect channel state information (CSI) available to optimize the beamformers. However, such an assumption cannot be satisfied in practice as the  CSI is always corrupted by inevitable errors. Such errors can lead to sub-optimal selection decisions of the beamformers, which can result in significant performance loss. The effect of imperfect CSI is more detrimental to FD compared to the half-duplex (HD) case, as the CSI errors in the SI channel could significantly degrade their performance. We remark that the only design to consider imperfect CSI in an IRS-FD system is available in \cite{ge2021robust}, limited to a simplified point-to-point MISO FD system.

 Despite the fruitful results obtained in the realm of IRS-FD systems, these studies assume perfect channel state information (CSI) availability for optimizing the beamformers. However, in practical scenarios, this assumption cannot be fulfilled due to the presence of inevitable errors that corrupt the CSI. Such errors can lead to suboptimal decisions regarding the selection of beamformers resulting in significant performance degradation. The impact of imperfect CSI is particularly pronounced in FD systems compared to their half-duplex (HD) counterparts \cite{sheemar2021game}, as errors in the CSI pertaining to the SI channel can substantially impair system performance. The consideration of imperfect CSI in the context of IRS-FD systems has been addressed solely in the study presented in \cite{ge2021robust}, albeit limited to a simplified point-to-point MISO FD system. This study considered the secrecy rate as an objective function. It is noteworthy that the problem of WSR maximization for the IRS-FD system under imperfect CSI remains unexplored. Furthermore, the state-of-the-art also lacked detailed analysis for the achievable WSR in IRS-FD systems under imperfect CSI.

\subsection{Main Contributions}

\begin{figure}[t]
    \centering
\includegraphics[width=7.8cm,height=6cm]{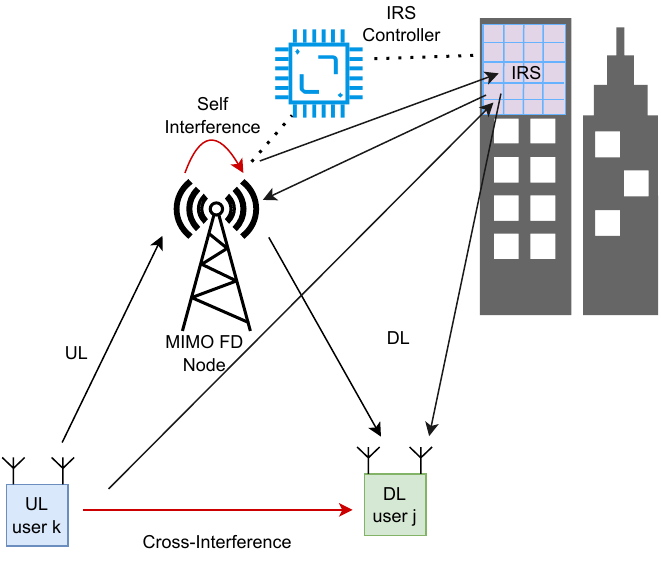}
    \caption{An IRS-FD system with multi-antenna UL and DL users.}
    \label{Scenario} \vspace{-2mm}
\end{figure}

\begin{itemize}
    \item Motivated by the aforementioned considerations, we delve into the problem of jointly optimizing the active and passive beamformers for an IRS-FD system with multi-antenna UL and DL user, shown in Fig. \ref{Scenario}. Recognizing the practical challenges in attaining perfect CSI, we address the case of imperfect CSI and adopt a statistically robust approach to design the beamformers for ergodic WSR maximization,  which pertain to the average WSR considering the CSI errors given the channel estimates. It is noteworthy that accounting for CSI errors in each link, alongside the presence of multi-antenna users, significantly amplifies the complexity of beamforming design for IRS-FD systems.

\item The Gaussian-Kronecker model \cite{rong2011robust, zeng2022joint} is adopted to characterize the CSI errors. By leveraging the statistical distribution information of these errors, we effectively analyze and quantify the impact of imperfect CSI on the IRS-FD system's performance. In our investigation, we take a rigorous analytical approach to derive both the lower and upper bounds of the achievable ergodic WSR, which enable us to assess the minimum and maximum capacity of the system in the presence of imperfect CSI, respectively.
%\item %We propose a novel beamforming design whose objective revolves around maximizing the ergodic WSR of the system under imperfect CSI conditions. The proposed approach is capable of handling uncertainty in CSI by exploiting the statistical distribution of the errors. Note that in contrast to the conventional approach without IRS, only digital beamfor

\item Subsequently, we formulate the problem of maximizing the ergodic WSR while simultaneously considering constraints on the total average sum-power at the base station and the unit-modulus phase-shift on the IRS response. To tackle this complex problem, we leverage the relationship between the ergodic WSR and the expected weighted minimum mean squared error (EWMMSE). This sophisticated approach takes into account the statistical properties of CSI errors during optimization, while considering the average mean squared error (MSE) covariance matrices. This transformation reduces the original problem into two distinct layers of sub-problems, which are iteratively solved. We note that the IRS also assists in passive SIC for FD systems. However, due to the imperfect CSI, its potential can be very limited. With the proposed robust approach, much higher levels of SIC can be achieved, resulting in a higher ergodic WSR for IRS-FD systems in the presence of CSI errors.

\item Finally, a comprehensive set of extensive simulation results is presented to verify and substantiate the effectiveness of the proposed robust joint beamforming design. The results corroborate the high accuracy of the derived analytical ergodic WSR approximation. Furthermore, the proposed statistically robust design is rigorously benchmarked against various schemes, including the robust IRS-assisted HD system. Based on the outcomes obtained, it is concluded that the proposed design significantly outperforms the benchmark schemes, solidifying its superiority and efficacy under imperfect CSI.
\end{itemize}

%We model the CSI errors with the Gaussian-Kronecker model \cite{rong2011robust,zeng2022joint}. Given the statistical distribution of the CSI errors, we first derive an analytical result for the upper and lower bounds for the ergodic achievable WSR under imperfect CSI. Then, we formulate the problem of ergodic WSR maximization under the total average sum-power and the unit-modulus constraint for the IRS phase response. 
%To solve such a problem, we resort to the link between the ergodic WSR and the expected weighted minimum mean squared error (EWMMSE). Such a method considers average mean squared error (MSE) covariance matrices based on the statistical distribution of the CSI errors while optimizing the variables involved in the optimization problem. The original ergodic WSR problem is reduced into two layers of sub-problems. The first layer tackles the active beamforming part by iteratively optimizing the digital beamformers, receive filter and the weight matrices by exploiting the convexity properties of each sub-problem. The second layer optimizes the passive beamforming part, which involves the IRS phase response. However, for the latter case, the optimization problem is still non-convex due to the unit-modulus constraint imposed on the phase response. To overcome this challenge, the majorization-minimization method \cite{sun2016majorization} is leveraged, which consists in reducing a challenging non-convex optimization problem into a series of simple and tractable problems.

 \emph{Paper Organization:} 
The rest of the paper is organized as follows. In Section \ref{sistema}, we first present the system model, model the CSI errors and formulate the problem of ergodic WSR maximization. In Sections \ref{Rate Analysis} and \ref{algorithmo}, we derivate the ergodic WSR and present a novel beamforming scheme that exhibits robustness based on the WMMSE. Finally, in Sections \ref{risultati} and \ref{Conc}, we present the numerical results and draw meaningful conclusions, respectively.

 \emph{Mathematical Notations:} Boldface lower and upper case characters denote vectors and matrices, respectively. $\mathbb{E}\{\cdot\}$, $\mbox{Tr}\{\cdot\}$, $\bmI$, denote expectation, trace, identity matrix, respectively, $\widetilde{\bmX}$ denotes the transmit covariance matrix $\widetilde{\bmX} = \bmX \bmX^H$. The superscripts $(\cdot)^T$ and $(\cdot)^H$ denote transpose
and conjugate-transpose (Hermitian) operators, respectively. The $\mbox{diag}(\bmx)$ denote a diagonal matrix with vector $\bmx$ on its main diagonal and $\odot$ denotes Hadamard product. Estimate of matrix $\bmX$ is denoted with $\hat{\bmX}$. $\mathbb{E}_{\bmX|\hat{\bmX}}$ denotes that expectation is taken with respect to $\bmX$, given its estimate $\hat{\bmX}$.
 
\section{System Model} \label{sistema}
Let $j$ and $k$ denote the multi-antenna DL and UL users served by the MIMO FD base station (BS), respectively, and
let $N_{j}$ and $M_{k}$ denote their number of receive and transmit antennas, respectively.
The FD BS is assumed to have $M_0$ transmit and $N_0$ receive antennas.  
We consider a multi-stream approach, and the number of data streams for the UL user $k$ and DL user $j$ are denoted as $u_k$ and $v_j$, respectively. Let $\bmU_k \in \mathbb{C}^{M_k \times u_k}$ and $\bmV_j \in \mathbb{C}^{M_0 \times v_j}$ denote the precoders for white unitary-variance data streams $\bms_{k}\in \mathbb{C}^{u_k \times 1}$ and $\bms_{j} \in \mathbb{C}^{v_j \times 1}$, respectively. We assume that the considered FD system is aided with one IRS of size $R \times C$.
Let $\bm{\theta} = [e^{i \theta_1},....,e^{i \theta_{RC}}]$ denote the vector containing the phase-shift response of its $RC$ elements, and let $\mathbf{\Theta} = \mbox{diag}(\bm{\theta})$ denote a diagonal matrix containing $\bm{\theta}$ on its main diagonal. Let $\bmn_0$ and 
$\bmn_j$ denote the noise vectors at the FD BS and DL user $j$, respectively, which are modelled as

\begin{equation}
    \bmn_0 = \mathcal{CN}(\bm0,\sigma_0^2 \bmI), \quad \bmn_j = \mathcal{CN}(\bm0,\sigma_j^2 \bmI),
\end{equation}
where  $\sigma_0^2$ and $\sigma_j^2$ denote the noise variances at the BS and the DL user $j$, respectively.

The channel responses from the UL user $k$ to the BS and from the BS to the DL user $j$ are denoted with  $\bmH_k \in \mathbb{C}^{N_0 \times M_k}$ and $\bmH_j \in \mathbb{C}^{N_j \times M_0}$, respectively. Let $\bmH_0\in  \mathbb{C}^{N_0 \times M_0}$ and $\bmH_{j,k} \in  \mathbb{C}^{N_j \times M_k}$ denote the SI channel response for the FD BS and cross-interference channel response between the UL user $k$ and the DL user $j$, respectively. The channel responses from the transmit antenna array of the FD BS to the IRS and from the IRS to the receive antenna array of the FD BS are denoted with $\bmH_{\theta,0} \in \mathbb{C}^{R C \times M_0}$
and $\bmH_{0,\theta} \in \mathbb{C}^{N_0 \times R C}$, respectively. Finally, $\bmH_{j,\theta} \in \mathbb{C}^{N_j \times RC}$ and $\bmH_{\theta,k} \in \mathbb{C}^{RC \times M_k}$ denote the channel responses from the IRS to the DL user $j$ and from the UL user $k$ to the IRS, respectively.

\subsection{Imperfect CSI Modelling}
%For the channel matrices described above, we assume that they have been estimated first, which can be for example acquired by adopting techniques like \cite{wang2020channel,nadeem2019intelligent,wei2021channel}. However, we remark that the CSI acquisition problem for the IRS-FD systems has not yet been investigated. Although different approaches can be adopted to 
%acquire CSI, such estimates are subject to errors which are inevitable in practice, leading to imperfect CSI. Note that imperfect CSI causes residual SI, interference and cross-interference in the FD systems, which can degrade their performance considerably if not considered in the optimization.

In the context of the aforementioned channel matrices, it is assumed that they have been initially estimated, potentially through techniques such as those proposed in \cite{wang2020channel,nadeem2019intelligent,wei2021channel}. However, it should be emphasized that the problem of CSI acquisition specifically for IRS-FD systems has not yet been thoroughly investigated. While various approaches can be employed for CSI acquisition, it is important to acknowledge that these estimates are prone to errors which are inevitable in practical scenarios resulting in imperfect CSI. It is crucial to account for the presence of imperfect CSI as it gives rise to residual SI and interference in IRS-FD systems, thereby significantly impacting their overall performance if not appropriately considered during the optimization process.

Let $\Delta\bmH_k$, $\Delta\bmH_j$, $\Delta\bmH_0$, $\Delta\bmH_{j,k}$, $\Delta\bmH_{\theta,0}$, $\Delta\bmH_{0,\theta}$, $\Delta\bmH_{j,\theta}$, and $ \Delta\bmH_{\theta,k}$ denote the estimation errors for the channel responses $\bmH_k$, $\bmH_j$, $\bmH_0$, $\bmH_{j,k}$, $\bmH_{\theta,0}$, $\bmH_{0,\theta}$, $\bmH_{j,\theta}$, and $\bmH_{\theta,k}$, respectively. The true CSI matrices can be written as a sum of the channel estimates and CSI errors as
\begin{equation} \label{channel_decomp}
    \begin{aligned}
     & \bmH_k  =  \hat{\bmH}_k + \Delta \bmH_k, \quad \quad \quad \; \bmH_j  = \hat{\bmH}_j + \Delta \bmH_j, \\
     & \bmH_0  = \hat{\bmH}_0 + \Delta \bmH_0, \quad \quad \quad \;  \hspace{0.7mm}
      \bmH_{j,k}= \hat{\bmH}_{j,k} + \Delta \bmH_{j,k}, \\
      &\bmH_{\theta,0} = \hat{\bmH}_{\theta,0} + \Delta \bmH_{\theta,0}, \;\quad \;
     \bmH_{0,\theta} = \hat{\bmH}_{0,\theta} + \Delta\bmH_{0,\theta}\\
    &  \bmH_{j,\theta} = \hat{\bmH}_{j,\theta} + \Delta\bmH_{j,\theta}, \quad \;\;\;
      \bmH_{\theta,k} = \hat{\bmH}_{\theta,k} + \Delta\bmH_{\theta,k},\\
    \end{aligned}      
\end{equation}
% \begin{equation} \label{channel_decomp}
%     \begin{cases} 
%       \bmH_k  =  \hat{\bmH}_k + \Delta \bmH_k, \; \bmH_j  = \hat{\bmH}_j + \Delta \bmH_j, \\
%      \bmH_0  = \hat{\bmH}_0 + \Delta \bmH_0, \;
%       \bmH_{j,k}= \hat{\bmH}_{j,k} + \Delta \bmH_{j,k}, \\
%      \bmH_{\theta,0} = \hat{\bmH}_{\theta,0} + \Delta \bmH_{\theta,0}, \;
%      \bmH_{0,\theta} = \hat{\bmH}_{0,\theta} + \Delta\bmH_{0,\theta},\\
%      \; \bmH_{j,\theta} = \hat{\bmH}_{j,\theta} + \Delta\bmH_{j,\theta}, \;
%       \bmH_{\theta,k} = \hat{\bmH}_{\theta,k} + \Delta\bmH_{\theta,k},\\
%     \end{cases}      
% \end{equation}
where the channel matrices of the form $\hat{\bmX}$ denote the channel estimates. To model the estimation errors, we adopt the Gaussian Kronecker model \cite{rong2011robust}, which dictates that

%\begin{equation} \label{CE_modelling}
%    \begin{aligned}
 %      &  \Delta \bmH_k = \mathcal{CN}(\bm0,\bmJ_k  \otimes \bmK_k ), \; \Delta \bmH_{\theta,0} = \mathcal{CN}(\bm0,\bmJ_{\theta,0} \otimes \bmK_{\theta,0} ), \\
  %  &  \Delta \bmH_0 = \mathcal{CN}(\bm0,\bmJ_0  \otimes \bmK_0 ), \;
   %   \Delta \bmH_{j,k} = \mathcal{CN}(\bm0,\bmJ_{j,k} \otimes \bmK_{j,k}), \\
    %  & \Delta \bmH_j = \mathcal{CN}(\bm0,\bmJ_j  \otimes \bmK_j), \; 
    % \Delta\bmH_{0,\theta} = \mathcal{CN}(\bm0,\bmJ_0  \otimes \bmK_{\theta}), \\
%    &  \Delta\bmH_{j,\theta} =\mathcal{CN}(\bm0,\bmJ_{j,\theta}  \otimes \bmK_{j,\theta} ), \;  \Delta\bmH_{\theta,k}= \mathcal{CN}(\bm0,\bmJ_{\theta,k}  \otimes \bmK_{\theta,k} ),\\
 %   \end{aligned}          
%\end{equation}

\begin{equation} \label{CE_modelling}
     \begin{cases}  
       \Delta \bmH_k = \mathcal{CN}\Big(\bm0,\bmJ_k  \otimes \bmK_k \Big),  \\ 
         \Delta \bmH_{\theta,0} = \mathcal{CN}\Big(\bm0,\bmJ_{\theta,0} \otimes \bmK_{\theta,0} \Big), \\
       \Delta \bmH_0 = \mathcal{CN}\Big(\bm0,\bmJ_0  \otimes \bmK_0 \Big),  \\
       \Delta \bmH_{j,k} = \mathcal{CN}\Big(\bm0,\bmJ_{j,k} \otimes \bmK_{j,k}\Big), \\
        \Delta \bmH_j = \mathcal{CN}\Big(\bm0,\bmJ_j  \otimes \bmK_j\Big),  \\
     \Delta\bmH_{0,\theta} = \mathcal{CN}\Big(\bm0,\bmJ_0  \otimes \bmK_{\theta}\Big), \\
       \Delta\bmH_{j,\theta} =\mathcal{CN}\Big(\bm0,\bmJ_{j,\theta}  \otimes \bmK_{j,\theta} \Big),  \\ \Delta\bmH_{\theta,k}= \mathcal{CN}\Big(\bm0,\bmJ_{\theta,k}  \otimes \bmK_{\theta,k} \Big),\\
  \end{cases}      
\end{equation}
where the matrices $\bmJ$ and $\bmK$ are the covariance matrices seen from the transmitter and receiver \cite{rong2011robust,zeng2022joint}, respectively. The estimation errors are uncorrelated with the estimated channel matrices \cite{yoo2006capacity}, and hence we have 

\begin{equation} \label{Channel_modelling_err}
    \begin{cases}
        \bmH_k = \mathcal{CN}\Big(\hat{\bmH}_k,\bmJ_k  \otimes \bmK_k\Big), \\ 
        \bmH_{\theta,0} = \mathcal{CN}\Big(\hat{\bmH}_{\theta,0},\bmJ_{\theta,0} \otimes \bmK_{\theta,0}\Big),  \\
      \bmH_0 = \mathcal{CN}\Big(\hat{\bmH}_0,\bmJ_0  \otimes \bmK_0\Big), \\
     \bmH_{j,k} = \mathcal{CN}\Big(\hat{\bmH}_{j,k},\bmJ_{j,k} \otimes \bmK_{j,k}\Big), \\
       \bmH_j = \mathcal{CN}\Big(\hat{\bmH}_j,\bmJ_j  \otimes \bmK_j\Big), \\
     \bmH_{0,\theta} = \mathcal{CN}\Big(\hat{\bmH}_{0,\theta},\bmJ_{0,\theta} \otimes \bmK_{0,\theta}\Big), \\
  \bmH_{j,\theta} =\mathcal{CN}\Big(\hat{\bmH}_{j,\theta},\bmJ_{j,\theta}  \otimes \bmK_{j,\theta}\Big),\\  \bmH_{\theta,k}= \mathcal{CN}\Big(\hat{\bmH}_{\theta,k},\bmJ_{\theta,k}  \otimes \bmK_{\theta,k}\Big).\\
    \end{cases}       
\end{equation}

Let $\overline{\bmH}_{0,k}, \overline{\bmH}_{0,0}, \overline{\bmH}_{j},\overline{\bmH}_{j,k}$ denote the effective channel responses affected by the estimation errors defined as  

\begin{subequations} \label{eff_channel}
\begin{equation}
   \overline{\bmH}_{k} = \Big(\hat{\bmH}_k + \Delta \bmH_k \Big) +  \Big(\hat{\bmH}_{0,\theta}  + \Delta \bmH_{0,\theta} \Big) \mathbf{\Theta} \Big(\hat{\bmH}_{\theta,k} + \Delta\bmH_{\theta,k}\Big), 
\end{equation} 
\begin{equation}
  \overline{\bmH}_{0} =  \Big(\hat{\bmH}_0 + \Delta \bmH_0\Big) + \Big(\hat{\bmH}_{0,\theta} + \Delta \bmH_{0,\theta} \Big) \mathbf{\Theta} \Big(\hat{\bmH}_{\theta,0} + \Delta\bmH_{\theta,0}\Big), 
\end{equation}
\begin{equation}   
\overline{\bmH}_{j}=\Big(\hat{\bmH}_j + \Delta \bmH_j\Big) + \Big(\hat{\bmH}_{j,\theta}  + \Delta \bmH_{j,\theta}\Big) \mathbf{\Theta} \Big(\hat{\bmH}_{\theta,0}  + \Delta\bmH_{\theta,0}\Big), 
\end{equation}
\begin{equation}
   \overline{\bmH}_{j,k} = \Big(\hat{\bmH}_{j,k} + \Delta \bmH_{j,k}\Big)  +  \Big(\hat{\bmH}_{j,\theta} +\Delta\bmH_{j,\theta}\Big) \mathbf{\Theta} \Big(\hat{\bmH}_{\theta,k} + \Delta  \bmH_{\theta,k} \Big). 
\end{equation}
\end{subequations}
Further, let $\bmy_k$ and $\bmy_j$ denote the signals received by the FD BS, from UL user $k$, and by the DL user $j$, respectively. By using \eqref{eff_channel}, they can be written as

\begin{subequations}
    
%\begin{figure*}[t]  
\begin{equation}  
\begin{aligned}
       \bmy_k = & \overline{\bmH}_k \bmU_k  \bms_{k} +   \overline{\bmH}_0 \bmV_j \bms_{j} + \bmn_0 ,
\end{aligned} \label{BS_side_CE}
\end{equation}
\begin{equation} 
\begin{aligned}
         \bmy_j = & \overline{\bmH}_j  \bmV_j  \bms_j  + \overline{\bmH}_{j,k}  \bmU_k  \bms_{k} + \bmn_j.
\end{aligned} \label{Rx_side_CE}
\end{equation}  
%\end{figure*}
\end{subequations}

\subsection{Problem Formulation}
 
In light of the corrupted CSI across all nodes, designing the beamformer and IRS phase response based solely on the knowledge of estimates $\hat{\bmH}_k, \hat{\bmH}_j, \hat{\bmH}_0, \hat{\bmH}_{j,k}, \hat{\bmH}_{\theta,0}, \hat{\bmH}_{0,\theta}, \hat{\bmH}_{j,\theta}, \hat{\bmH}_{\theta,k}$ can result in substantial performance degradation attributable to the inherent mismatch.

%In this work, we aim to maximize the ergodic WSR of the IRS-FD system under imperfect CSI given the CSI errors statistics and the channel estimates. Let $\mathcal{R} = \mathcal{R}_k + \mathcal{R}_j$ denote the WSR of the system in the case of perfect CSI, with $\mathcal{R}_k$ and $\mathcal{R}_j$ denoting the weighted rate of the users $k$ and $j$, respectively. The ergodic WSR of the system, i.e., the average WSR with respect to the CSI errors, can be written as $\mathbb{E}_{\bmH|\hat{\bmH}}[\mathcal{R}]$. However, such problem is difficult to solve and can be simplified with the Jensen equality, which allows to move the expectation operator on the arguments of the $\mbox{log}(\cdot)$ as  $\mathbb{E}_{\bmH|\hat{\bmH}}[\mathcal{R}] \geq \mathcal{R}(\mathbb{E}_{\bmH|\hat{\bmH}})$ \cite{negro2012sum}. For ergodic WSR $\mathcal{R}(\mathbb{E}_{\bmH|\hat{\bmH}})$, the maximization problem can be formally stated as

In this endeavor, our objective is to maximize the ergodic WSR of the MIMO IRS-FD system considering the presence of imperfect CSI along with the statistics of CSI errors and channel estimates. Let $\mathcal{R} = \mathcal{R}_k + \mathcal{R}_j$ represent the WSR of the system under perfect CSI, where $\mathcal{R}_k$ and $\mathcal{R}_j$ denote the weighted rates of users $k$ and $j$, respectively. The ergodic WSR, which captures the average WSR considering the CSI errors, can be expressed as $\mathbb{E}_{\bmH|\hat{\bmH}}[\mathcal{R}]$. However, solving this problem directly poses challenges. To address this, the Jensen's inequality is employed, enabling the relocation of the expectation operator onto the arguments of the logarithm, yielding $\mathbb{E}_{\bmH|\hat{\bmH}}[\mathcal{R}] \geq \mathcal{R}(\mathbb{E}_{\bmH|\hat{\bmH}})$ \cite{negro2012sum}, where $\mathcal{R}(\mathbb{E}_{\bmH|\hat{\bmH}})$ emphasize that the rate is evaluated with the
expectation taken with respect to the channel, given the estimates. Formally, the maximization problem for the ergodic WSR $\mathcal{R}(\mathbb{E}_{\bmH|\hat{\bmH}})$ can be formulated as follows:

\begin{subequations}\label{WSR_problem}
\begin{equation}
\underset{\substack{\bmV_j,\bmU_k},\mathbf{\Theta}}{\max} \quad  \mathcal{R}_k(\mathbb{E}_{\bmH|\hat{\bmH}}) + \mathcal{R}_j(\mathbb{E}_{\bmH|\hat{\bmH}})
\end{equation} \label{WSR}
\begin{equation}
\text{s.t.}  \quad\mbox{Tr}\Big( \bmU_k \bmU_k^H \Big) 	\preceq \alpha_k,  \label{c1}
\end{equation}
\begin{equation}
\quad \quad    \mbox{Tr} \Big( \bmV_j  \bmV_j^H \Big) \leq  \alpha_0, \label{c2}
\end{equation}
\begin{equation} \label{c3}
 \quad 
  \quad   \quad  \Big|\bm{\theta}(i)\Big|=1,  \quad \forall i.
\end{equation}
\end{subequations} 
The statistical expectation is taken with respect to the CSI, with the distribution given in \eqref{Channel_modelling_err}. As the precise CSI remains unknown, the constraints \eqref{c1}-\eqref{c2} denote the average power constraint in UL and DL, given the channel estimates $\hat{\bmH}$, and \eqref{c3} denotes the unit-modulus constraint imposed on the IRS phase-response.

\section{Ergodic WSR Analysis with Imperfect CSI} \label{Rate Analysis}

%The WSR of MIMO FD system with perfect CSI and without the IRS can be written as  \cite{cirik2015weighted}
%\begin{equation} \label{upper}
%\begin{aligned}
% \mathcal{R} = &   w_k \mbox{ln}\Big[\mbox{det}\Big( \bmU_k^H \bmH_k^H \bmR_{\overline{k}}^{-1} \bmH_k \bmU_k \Big) \Big] \\& +  w_j \mbox{ln}\Big[\mbox{det}\Big( \bmV_j^H \bmH_j^H \bmR_{\overline{j}}^{-1} \bmH_j \bmV_j \Big) \Big],
%\end{aligned}
%\end{equation}
%where $\bmR_{\overline{k}}$ and $\bmR_{\overline{j}}$ denote the interference plus noise covariance matrices, function of the perfect CSI.

In the subsequent analysis, we derive the expression for the ergodic WSR $\mathcal{R}(\mathbb{E}_{\bmH|\hat{\bmH}})$ considering the presence of imperfect CSI in the MIMO IRS-FD system. We define $\widetilde{\bmU}_k = \bmU_k \bmU_k^H$ and $\widetilde{\bmV}_j = \bmV_j \bmV_j^H$ as the transmit covariance matrices for UL user $k$ and DL user $j$, respectively. When imperfect CSI is taken into account, the received signal plus interference and noise covariance matrices, denoted as $\bmR_{k}$ and $\bmR_{j}$, encompassing both the CSI errors and the IRS phase response $\mathbf{\Theta}$, can be expressed as follows:

\begin{subequations} 
    \begin{equation}  \label{Ul_user_cov} 
    \begin{aligned}
        \bmR_{k} = & \overline{\bmH}_k \widetilde{\bmU}_k \overline{\bmH}_k^H + \overline{\bmH}_0 \widetilde{\bmV}_j \overline{\bmH}_0^H + \sigma_0^2 \bmI.
    \end{aligned}
    \end{equation}
     \begin{equation} \label{Dl_user_cov} 
    \begin{aligned}
        \bmR_{j} = & \overline{\bmH}_j \widetilde{\bmV}_j \overline{\bmH}_j^H + \overline{\bmH}_{j,k} \widetilde{\bmU}_k   \overline{\bmH}_{j,k}^H   + \sigma_j^2 \bmI.
    \end{aligned}
    \end{equation}
\end{subequations}
The interference plus noise covariance matrices can be obtained as $\bmR_{\overline{k}} =  \bmR_{k} - \bmS_k, \bmR_{\overline{j}} =  \bmR_{j} - \bmS_j$, with $\bmS_k$ and $\bmS_j$ denoting the useful received signal covariance part.

\newtheorem{theorem}{Theorem}
\begin{theorem} \label{theorem_1}
Given the statistical distribution of the CSI errors \eqref{CE_modelling} and the channel estimates, the ergodic WSR $\mathcal{R}(\mathbb{E}_{\bmH|\hat{\bmH}})$ of an MIMO IRS-FD system under imperfect CSI can be approximated as
     
\begin{equation} \label{ach_rate_hat}
\begin{aligned}
\mathcal{R}(\mathbb{E}_{\bmH|\hat{\bmH}}) &=    w_k \mbox{ln}\Big[\mbox{det}\Big(\bmI +  \bmU_k^H \Big(\hat{\bmH}_{k} +  \hat{\bmH}_{0,\theta} \mathbf{\Theta} \hat{\bmH}_{\theta,k}\Big)^H \mathbf{\Sigma}_{\overline{k}}^{-1} \\& \quad \quad\quad \quad \quad\quad \Big(\hat{\bmH}_{k} +  \hat{\bmH}_{0,\theta} \mathbf{\Theta} \hat{\bmH}_{\theta,k}\Big)  \bmU_k \Big) \Big] \\& +   w_j \mbox{ln}\Big[\mbox{det}\Big(\bmI +  \bmV_j^H \Big(\hat{\bmH}_{j} + \hat{\bmH}_{j,\theta}   \mathbf{\Theta} \hat{\bmH}_{\theta,0}\Big)^H \mathbf{\Sigma}_{\overline{j}}^{-1}\\& \quad \quad\quad \quad \quad\quad \Big(\hat{\bmH}_{j} + \hat{\bmH}_{j,\theta}   \mathbf{\Theta} \hat{\bmH}_{\theta,0}\Big) \bmV_j \Big) \Big],
\end{aligned}
\end{equation}
where $\mathbf{\Sigma}_{\overline{k}}$ and $\mathbf{\Sigma}_{\overline{j}}$ are given as in \eqref{cov_true}, at the top of the next page and $w_k$ and $w_j$ denote the weights.
\end{theorem}

\begin{figure*}[!t] 
\begin{subequations} \label{cov_true}
\begin{equation}  \label{cov_UL_user_k}
\begin{aligned}
    \mathbf{\Sigma}_{\overline{k}} &=   \mbox{Tr}\Big(\widetilde{\bmU}_k \bmJ_k^T\Big) \bmK_k      + \hat{\bmH}_{0,\theta} \mathbf{\Theta} \mbox{Tr}\Big(\widetilde{\bmU}_k \bmJ_{\theta,k}^T\Big) \bmK_{\theta,k} \mathbf{\Theta}^H \hat{\bmH}_{0,\theta}^H  + \mbox{Tr}\Big(\mathbf{\Theta} \hat{\bmH}_{\theta,k} \widetilde{\bmU}_k \hat{\bmH}_{\theta,k} \mathbf{\Theta}^H \bmJ_{0,\theta}^T\Big) \bmK_{0,\theta}   + \mbox{Tr}\Big(\widetilde{\bmU}_k \bmJ_{\theta,k}^T\Big)  \\& \times\mbox{Tr}\Big(\mathbf{\Theta} \bmK_{\theta,k} \mathbf{\Theta}^H \bmJ_{0,\theta}^T\Big) \bmK_{0,\theta} 
    + \hat{\bmH}_0 \widetilde{\bmV}_j \hat{\bmH}_0^H  + \hat{\bmH}_0 \widetilde{\bmV}_j \bmH_{\theta,0}^H \mathbf{\Theta}^H \hat{\bmH}_{0,\theta} + \mbox{Tr}\Big(\widetilde{\bmV}_j \bmJ_0^T\Big) \bmK_0^T + \hat{\bmH}_{0,\theta} \mathbf{\Theta} \hat{\bmH}_{\theta,0} \widetilde{\bmV}_j \hat{\bmH}_0^H \\& + \hat{\bmH}_{0,\theta} \mathbf{\Theta} \bmH_{\theta,0} \widetilde{\bmV}_j \bmH_{\theta,0}^H \mathbf{\Theta}^H \hat{\bmH}_{0,\theta}^H + \hat{\bmH}_{0,\theta} \mathbf{\Theta} \mbox{Tr}\Big(\widetilde{\bmV}_j \bmJ_{\theta,0}^T\Big) \bmK_{\theta,0} \mathbf{\Theta}^H \hat{\bmH}_{0,\theta}^H   + \mbox{Tr}\Big(\mathbf{\Theta} \hat{\bmH}_{\theta,0} \widetilde{\bmV}_j \hat{\bmH}_{\theta,0}^H \mathbf{\Theta}^H \bmJ_{0,\theta}^T\Big) \bmK_{0,\theta}  \\& + \mbox{Tr}\Big(\widetilde{\bmV}_j \bmJ_{\theta,0}^T\Big) \mbox{Tr}\Big(\mathbf{\Theta} \bmK_{\theta,0} \mathbf{\Theta}^H \bmJ_{0,\theta}^T\Big) \bmK_{0,\theta} + \sigma_0^2 \bmI,
\end{aligned}   
\end{equation}
\begin{equation} 
    \begin{aligned}
       \mathbf{\Sigma}_{\overline{j}} &=     \mbox{Tr}\Big(\widetilde{\bmV}_j \bmJ_j^T\Big) \bmK_j  + \hat{\bmH}_{j,\theta} \mathbf{\Theta} \mbox{Tr}\Big(\widetilde{\bmV}_j \bmJ_{\theta,0}^T\Big) \bmK_{\theta,0} \mathbf{\Theta}^H \hat{\bmH}_{j,\theta}^H  + \mbox{Tr}\Big( \mathbf{\Theta} \hat{\bmH}_{\theta,0} \widetilde{\bmV}_j \hat{\bmH}_{\theta,0}^H \mathbf{\Theta}^H \bmJ_{j,\theta}^T \Big) \bmK_{j,\theta} + \mbox{Tr}\Big(\widetilde{\bmV}_j \bmJ_{\theta,0}^T\Big) \\&\times\mbox{Tr}\Big(\mathbf{\Theta} \bmK_{\theta,0} \mathbf{\Theta}^H \bmJ_{j,\theta}^T\Big) \bmK_{j,\theta}  + \hat{\bmH}_{j,k} \widetilde{\bmU}_k \hat{\bmH}_{j,k}^H  +\hat{\bmH}_{j,k} \widetilde{\bmU}_k \hat{\bmH}_{\theta,k}^H \mathbf{\Theta}^H \hat{\bmH}_{j,\theta}^H + \mbox{Tr}\Big(\widetilde{\bmU}_k \bmJ_{j,k}^T\Big) \bmK_{j,k}   + \hat{\bmH}_{j,\theta} \mathbf{\Theta} \bmH_{\theta,k} \widetilde{\bmU}_k \hat{\bmH}_{j,k}\\& + \hat{\bmH}_{j,\theta} \mathbf{\Theta} \bmH_{\theta,k} \widetilde{\bmU}_k \hat{\bmH}_{\theta,k}^H \mathbf{\Theta}^H \hat{\bmH}_{j,\theta}^H   + \hat{\bmH}_{j,\theta} \mathbf{\Theta} \mbox{Tr}\Big(\widetilde{\bmU}_k \bmJ_{\theta,k}^T\Big) \bmK_{\theta,k} \mathbf{\Theta}^H \hat{\bmH}_{j,\theta}^H  + \mbox{Tr}\Big(\mathbf{\Theta} \hat{\bmH}_{\theta,k} \widetilde{\bmU}_k \hat{\bmH}_{\theta,k}^H \mathbf{\Theta}^H \bmJ_{j,\theta}^T\Big) \bmK_{j,\theta} \\& + \mbox{Tr}\Big(\widetilde{\bmU}_k \bmJ_{\theta,k}^T \Big) \mbox{Tr}\Big(\mathbf{\Theta} \bmK_{\theta,k}  \mathbf{\Theta}^H \bmJ_{j,\theta}^T \Big)  \bmK_{j,\theta} + \sigma_j^2 \bmI. 
\end{aligned}
\end{equation}  
\end{subequations} \hrulefill
\end{figure*}

\begin{proof}
We prove the result for UL user $k$, and a similar reasoning can be carried out for the DL user $j$. The total received covariance matrix $\bmR_{k}$ in \eqref{Ul_user_cov}, given the CSI estimates and the CSI errors, can be written as

\begin{equation} \label{ach_rate_k}
\begin{aligned}
    \mathbf{R}_{k} &=  \hat{\bmH}_{k} \widetilde{\bmU}_k \hat{\bmH}_{k}^H  + \Delta \bmH_k \widetilde{\bmU}_k \Delta \bmH_k^H +  \hat{\bmH}_{k} \widetilde{\bmU}_k \hat{\bmH}_{\theta,k}^H \mathbf{\Theta}^H \hat{\bmH}_{0,\theta}^H \\&  +   \hat{\bmH}_{0,\theta} \mathbf{\Theta} \hat{\bmH}_{\theta,k} \widetilde{\bmU}_k \hat{\bmH}_k^H    +  \hat{\bmH}_{0,\theta} \mathbf{\Theta} \hat{\bmH}_{\theta,k} \widetilde{\bmU}_k \hat{\bmH}_{\theta,k}^H \mathbf{\Theta}^H \hat{\bmH}_{0,\theta}^H \\& + 
     \hat{\bmH}_{0,\theta} \mathbf{\Theta}  \hat{\bmH}_{\theta,k} \widetilde{\bmU}_k   \hat{\bmH}_{\theta,k}^H \mathbf{\Theta}^H \hat{\bmH}_{0,\theta}^H 
    \\& +  \hat{\bmH}_{0,\theta} \mathbf{\Theta} \Delta \hat{\bmH}_{\theta,k} \widetilde{\bmU}_k \Delta \hat{\bmH}_{\theta,k}^H \mathbf{\Theta}^H \hat{\bmH}_{0,\theta}^H  \\&
    +  \Delta \hat{\bmH}_{0,\theta} \mathbf{\Theta} \Delta \hat{\bmH}_{\theta,k} \widetilde{\bmU}_k   \Delta \hat{\bmH}_{\theta,k}^H \mathbf{\Theta}^H  \Delta \hat{\bmH}_{0,\theta}^H 
    \\&
    +  \Delta \hat{\bmH}_{0,\theta} \mathbf{\Theta} \hat{\bmH}_{\theta,k} \widetilde{\bmU}_k \hat{\bmH}_{\theta,k}^H \mathbf{\Theta}^H \Delta \hat{\bmH}_{0,\theta}^H   \\&  +  \hat{\bmH}_0 \widetilde{\bmV}_j \hat{\bmH}_0^H    + \Delta\bmH_0 \widetilde{\bmV}_j \Delta \bmH_0^H +   \hat{\bmH}_0  \widetilde{\bmV}_j \bmH_{\theta,0}^H \mathbf{\Theta}^H \hat{\bmH}_{0,\theta}^H \\&
     +  \hat{\bmH}_{0,\theta}  \mathbf{\Theta} \bmH_{\theta,0}  \widetilde{\bmV}_j \hat{\bmH}_0^H   +  \bmH_{0,\theta} \mathbf{\Theta}  \hat{\bmH}_{\theta,0}  \widetilde{\bmV}_j \hat{\bmH}_{\theta,0}^H \mathbf{\Theta}^H \bmH_{0,\theta}^H  \\& +  \hat{\bmH}_{0,\theta} \mathbf{\Theta} \Delta \bmH_{\theta,0}  \widetilde{\bmV}_j \Delta \bmH_{\theta,0}^H \mathbf{\Theta}^H \hat{\bmH}_{0,\theta}^H  \\& +  \Delta \bmH_{0,\theta} \mathbf{\Theta} \hat{\bmH}_{\theta,0}   \widetilde{\bmV}_j \hat{\bmH}_{\theta,0}^H \mathbf{\Theta}^H \Delta \bmH_{0,\theta}^H   \\& +  \Delta \bmH_{0,\theta} \mathbf{\Theta} \Delta \bmH_{\theta,0}   \widetilde{\bmV}_j \Delta \bmH_{\theta,0}^H \mathbf{\Theta}^H \Delta \bmH_{0,\theta}^H  + \bmL,
\end{aligned}
\end{equation} 
where $\bmL$ includes terms which are linear in the CSI errors.
Let $\mathbf{\Sigma}_k = \mathbb{E}_{\bmH|\hat{\bmH}}[\bmR_{k}]$ denote the expected signal plus interference plus noise covariance matrix, where the expectation is carried out with respect to the channel responses, including CSI errors. Consider the result given in \cite[Lemma~1]{rong2011robust} for imperfect CSI under the Gaussian-Kronecker model stating that: For $\bmH\sim \mathcal{CN}(\hat{\bmH},\bmJ \otimes \bmK) $, there is $\mathbb{E}[\bmH \bmX \bmH^H ] = \hat{\bmH
} \bmX \hat{\bmH}^H + \mbox{Tr}(\bmX \bmJ^T) \bmK  $ and  $\mathbb{E}[\bmH^H \bmX \bmH] = \hat{\bmH
}^H \bmX \hat{\bmH} + \mbox{Tr}(\bmK\bmX) \bmJ^T$. For the MIMO IRS-FD system, we first subtract the useful signal part from $\mathbf{\Sigma}_{k}$ which depends only on the estimated channel responses and then apply the result on each term of the expected interference plus noise covariance matrix $\mathbf{\Sigma}_{\overline{k}}$. Ignoring the terms which are linear in the CSI errors, and by applying the result above, it can be shown that the ergodic WSR of the UL user $k$ under imperfect CSI has the interference plus noise covariance matrix structure $\mathbf{\Sigma}_{\overline{k}}$ given in \eqref{cov_UL_user_k}.
\end{proof}
 
It will be shown in Section \ref{risultati} that the aforementioned approximation of the ergodic rate achieves a high level of accuracy. It is important to note that \eqref{ach_rate_hat} serves as a lower bound on the achievable WSR for an IRS-FD system operating under imperfect CSI conditions. Therefore, maximizing \eqref{ach_rate_hat} is equivalent to maximizing the worst-case WSR. Conversely, an upper bound on the achievable ergodic WSR for the IRS-FD system is given by considering the ideal CSI scenario, wherein the CSI error variance for all channels tends to zero.

\section{Robust Joint Active and Passive Beamforming Via EWMMSE } \label{algorithmo}

The problem of maximizing the ergodic WSR as formulated in \eqref{WSR_problem} is inherently non-convex due to the presence of interference. However, in the ideal scenario of perfect CSI, an equivalent problem formulation known as weighted minimum mean squared error (WMMSE) can be employed, leveraging the established relationship between WSR and MSE \cite{christensen2008weighted}. In the case of imperfect CSI, the WMMSE problem can be further transformed into the framework of EWMMSE, which incorporates the expectation with respect to the CSI errors \cite{joudeh2016sum,negro2012sum}. This approach involves considering the average MSE covariance matrices and we adopt this methodology throughout the paper.

\subsection{Digital Combining}
Assume that the FD BS for UL user $k$ and the DL user $j$  apply the combiners $\bmF_k$ and $\bmF_j$ to estimate their data streams as
\begin{equation} \label{combiners}
    \hat{\bms}_k = \bmF_k \bmy_k, \quad  \hat{\bms}_j = \bmF_j \bmy_j.
\end{equation}
Given \eqref{combiners}, let $\bmE_{\tilde{k}}$ and $\bmE_{\tilde{j}}$ denote the MSE error matrices for instantaneous CSI for $k$-th UL user and $j$-th DL user, respectively, which can be written as 
\begin{subequations}
    \begin{equation}
         \bmE_{\tilde{k}} =  \mathbb{E}\Big[(\bmF_k \bmy_k - \hat{\bms}_k) (\bmF_k \bmy_k - \hat{\bms}_k)^H\Big],
    \end{equation}
    \begin{equation}
         \bmE_{\tilde{j}}  =  \mathbb{E}\Big[(\bmF_j \bmy_j - \hat{\bms}_j) (\bmF_j \bmy_j - \hat{\bms}_j)^H\Big]. 
    \end{equation}
\end{subequations}

Let $\bmQ_k, \bmT_j, \bmT_{0}$, and $\bmQ_{j,k}$ denote the matrices defined as
\begin{subequations} \label{aux_Q}
    \begin{equation}
        \bmQ_k  =  \mathbb{E}_{\bmH|\hat{\bmH}}\Big(\overline{\bmH}_k \widetilde{\bmU}_k \overline{\bmH}_k^H\Big), \;  
         \bmT_j =  \mathbb{E}_{\bmH|\hat{\bmH}}\Big(\overline{\bmH}_j \widetilde{\bmV}_j \overline{\bmH}_j^H\Big),
    \end{equation}
    \begin{equation}
       \bmT_{0} = \mathbb{E}_{\bmH|\hat{\bmH}}\Big(\overline{\bmH}_0 \widetilde{\bmV}_j \overline{\bmH}_0^H\Big),  \;
        \bmQ_{j,k} =   \mathbb{E}_{\bmH|\hat{\bmH}}\Big(\overline{\bmH}_{j,k}  \widetilde{\bmU}_k \overline{\bmH}_{j,k}^H\Big),
    \end{equation}
\end{subequations}
which are given in Appendix \ref{aux-matrices}. The optimization of the combiners can be based on the expected MSE (EMSE) covariance matrices, given below
\begin{subequations} \label{error_cov}
    \begin{equation}
    \begin{aligned}
        \bmE_k & =   \mathbb{E}_{\bmH|\hat{\bmH}}\Big[\bmE_{\tilde{k}}\Big] \\ &= \bmF_k \bmQ_k \bmF_k^H - \bmF_k 
        (\hat{\bmH}_k + \hat{\bmH}_{0,\theta} \mathbf{\Theta} \hat{\bmH}_{\theta,k})  \bmU_k  
         + \bmF_k \bmT_{0} \bmF_k^H  \\& \quad + \sigma_0^2 \bmF_k \bmF_k^H - \bmU_k^H \hat{\bmH}_k^H \bmF_k^H   - \bmU_k^H \hat{\bmH}_{\theta,k}^H  \mathbf{\Theta}^H \hat{\bmH}_{0,\theta}^H \bmF_k^H - \bmI,
    \end{aligned}
    \end{equation}
        \begin{equation}
    \begin{aligned}
        \bmE_j  & =   \mathbb{E}_{\bmH|\hat{\bmH}}\Big[\bmE_{\tilde{j}}\Big] \\ &=
        \bmF_j  \bmT_j \bmF_j^H - \bmF_j (\hat{\bmH}_j + 
        \hat{\bmH}_{j,\theta}  \mathbf{\Theta} \hat{\bmH}_{\theta,0} )  \bmV_j 
      + \bmF_j  \bmQ_{j,k} \bmF_j^H \\& \quad  + \sigma_j^2 \bmF_j \bmF_j^H  - \bmV_j^H  \hat{\bmH}_j^H \bmF_j^H  -
       \bmV_j^H\hat{\bmH}_{\theta,0}^H \mathbf{\Theta}^H  \hat{\bmH}_{j,\theta}^H   \bmF_j^H - \bmI.
    \end{aligned}
    \end{equation}
\end{subequations}
The optimization problem of the combiners can be stated as the minimization of the error covariance matrices \eqref{error_cov} as

\begin{equation} \label{comb_solution}
 \underset{\substack{\bmF_l,\bmF_r}}{\min} \quad  \mbox{Tr}\Big(\bmE_k\Big) +  \mbox{Tr}\Big(\bmE_l\Big).
\end{equation}
 By solving \eqref{comb_solution}, we get the following optimal combiners 

\begin{subequations} \label{MMSE_comb}
\begin{equation}
    \bmF_k = \bmU_k^H \Big(\hat{\bmH}_k^H + \hat{\bmH}_{\theta,k}^H \mathbf{\Theta}^H \hat{\bmH}_{0,\theta}^H \Big) \Big( \bmQ_k +  \bmT_{0} + \sigma_0^2 \bmI \Big)^{-1},
\end{equation}
\begin{equation}
    \bmF_j = \bmV_j^H \big(\hat{\bmH}_j^H + \hat{\bmH}_{\theta,0}^H \mathbf{\Theta}^H \hat{\bmH}_{j,\theta}^H \big) \big(  \bmT_j  +\bmQ_{j,k} + \sigma_j^2 \bmI \big)^{-1}.
\end{equation}
\end{subequations}

%By plugging the expected MMSE combiners in the average MSE covariance matrices, it could be easily shown that the 

%\begin{subequations} \label{error_cov_final}
 %   \begin{equation}
  %  \begin{aligned}
   %     \bmE_k & = \Big( \bmI +   \bmU_k^H \Big(\hat{\bmH}_{k} +  \hat{\bmH}_{0,\theta} \mathbf{\Theta} \hat{\bmH}_{\theta,k}\Big)^H \mathbf{\Sigma}_{\overline{k}}^{-1} \\& \quad \quad\quad \quad \quad\quad \Big(\hat{\bmH}_{k} +  \hat{\bmH}_{0,\theta} \mathbf{\Theta} \hat{\bmH}_{\theta,k}\Big)  \bmU_k \Big)^{-1}, 
    %\end{aligned}
    %\end{equation}
    %   \begin{equation}
    %\begin{aligned}
     %   \bmE_j  & =   \Big(\bmI +  \bmV_j^H \Big(\hat{\bmH}_{j} + \hat{\bmH}_{j,\theta}   \mathbf{\Theta} \hat{\bmH}_{\theta,0}\Big)^H \mathbf{\Sigma}_{\overline{j}}^{-1}\\& \quad \quad\quad \quad \quad\quad \Big(\hat{\bmH}_{j} + \hat{\bmH}_{j,\theta}   \mathbf{\Theta} \hat{\bmH}_{\theta,0}\Big) \bmV_j \Big)^{-1}.
    %\end{aligned}
    %\end{equation}
%\end{subequations}
%Therefore, the achievable WSR under the imperfect CSI is related to the WMMSE as

%\begin{subequations}
  %  \begin{equation}
 %       \mathcal{R}_k(\mathbb{E}_{\bmH|\hat{\bmH}}) = \mbox{log}\Big[\mbox{det}\Big(\bmE_k^{-1}\Big)\Big],
 %   \end{equation}
 %    \begin{equation}
%\mathcal{R}_j(\mathbb{E}_{\bmH|\hat{\bmH}})= \mbox{log}\Big[\mbox{det}\Big(\bmE_j^{-1}\Big)\Big].
%    \end{equation}
%\end{subequations}

\subsection{Active Digital Beamforming Under Imperfect CSI}

Given the optimal combiners, the EWMMSE problem with respect to the digital beamformers under the average total sum-power constraint and given the IRS phase response $\mathbf{\Theta}$ fixed, can be formally stated as 

\begin{subequations}\label{dig_BF_problem}
\begin{equation}
     \underset{\substack{\bmV_j,\bmU_k}}{\min} \quad  \mbox{Tr}\Big(\bmW_k \bmE_k\Big) +   \mbox{Tr}\Big( \bmW_j \bmE_j\Big),
\end{equation} \label{WSR}
\begin{equation}
  \text{s.t.} \hspace{2mm}   \mbox{Tr}\Big( \bmU_k \bmU_k^H \Big) 	\preceq \alpha_k, \label{c1_2}
\end{equation}
\begin{equation}
\quad \;\;  \mbox{Tr} \Big(  \bmV_j  \bmV_j^H \Big) \leq  \alpha_0, \label{c2-2}
\end{equation}
\end{subequations}
where $\bmW_i$ is a constant weight matrix associated with node $i$. Problem \eqref{dig_BF_problem} and \eqref{WSR_problem} are equivalent if their gradients are the same, which can be ensured by selecting the weight matrices as follows:
\begin{equation} \label{W_matrix}
        \bmW_k = \frac{w_k}{\mbox{ln}\; 2} \Big( \bmE_k \Big)^{-1}, \quad
        \bmW_j = \frac{w_j}{\mbox{ln}\;2} \Big( \bmE_j \Big)^{-1}.
\end{equation}
The equivalence between the two problems can be demonstrated by following a similar proof as provided in \cite[Appendix A]{cirik2015weighted}, but adapted for the case of expected MSE under imperfect CSI.

 To optimize the digital beamformers $\bmV_j$ and $\bmU_k$, we calculate the partial derivative of the Lagrangian function of \eqref{dig_BF_problem} with respect to their conjugate. This calculation yields the following optimal beamformers.
\begin{figure*}
\begin{subequations}
    \begin{equation} \label{X_k}
    \begin{aligned}
        \bmX_{k} = & \Big(\hat{\bmH}_k^H + \hat{\bmH}_{\theta,k}^H \mathbf{\Theta}^H \hat{\bmH}_{0,\theta}^H\Big) \bmF_k^H \bmW_k \bmF_k \hat{\bmH}_k + \mbox{Tr}\Big(\bmK_k \bmF_k^H \bmW_k \bmF_k\Big)  \bmJ_k^T + \hat{\bmH}_{\theta,k}^H \mathbf{\Theta}^H \hat{\bmH}_{0,\theta}^H \bmF_k^H \bmW_k \bmF_k \hat{\bmH}_{0,\theta} \mathbf{\Theta} \hat{\bmH}_{\theta,k} \\& + \bmH_k^H \bmF_k^H \bmW_k \bmF_k \hat{\bmH}_{0,\theta} \mathbf{\Theta} \hat{\bmH}_{\theta,k} +
        \mbox{Tr}\Big(\bmK_{\theta,k} \mathbf{\Theta}^H \hat{\bmH}_{0,\theta}^H \bmF_k^H \bmW_k \bmF_k \hat{\bmH}_{0,\theta} \mathbf{\Theta} \Big)  \bmJ_{\theta,k}^T
         + \hat{\bmH}_{\theta,k}^H \mathbf{\Theta}^H
         \mbox{Tr}( \bmK_{0,\theta}\bmF_k^H \bmW_k \bmF_k )  \bmJ_{0,\theta}^T \mathbf{\Theta} \hat{\bmH}_{\theta,k}   \\&
        + \hat{\bmH}_{j,k}^H \bmF_j^H \bmW_j \bmF_j \hat{\bmH}_{j,\theta} \mathbf{\Theta} \hat{\bmH}_{\theta,k} 
            + \Big(\hat{\bmH}_{j,k}^H + \hat{\bmH}_{\theta,k}^H \mathbf{\Theta}^H \hat{\bmH}_{j,\theta}^H \Big)\bmF_j^H \bmW_j \bmF_j \hat{\bmH}_{j,k}  
           + \mbox{Tr}\Big(\bmK_{j,k} \bmF_j^H \bmW_j \bmF_j \Big)  \bmJ_{j,k}^T
           \\& + \mbox{Tr}\Big(\bmK_{0,\theta} \bmF_k^H \bmW_k \bmF_k) \mbox{Tr}(\bmK_{\theta,k} \mathbf{\Theta}^H \bmJ_{0,\theta}^T \mathbf{\Theta}\Big) \bmJ_{\theta,k}^T  + \hat{\bmH}_{\theta,k}^H \mathbf{\Theta}^H \hat{\bmH}_{j,\theta}^H \bmF_j^H \bmW_j \bmF_j \hat{\bmH}_{j,\theta} \mathbf{\Theta} \hat{\bmH}_{\theta,k} \\&+ 
           \mbox{Tr}\Big(\bmK_{\theta,k} \mathbf{\Theta}^H \hat{\bmH}_{j,\theta}^H \bmF_j^H \bmW_j \bmF_j \hat{\bmH}_{j,\theta} \mathbf{\Theta} \Big) \bmJ_{\theta,k}^T  
            + \hat{\bmH}_{\theta,k}^H \mathbf{\Theta}^H 
            \mbox{Tr}\Big(\bmK_{j,\theta} \bmF_j^H \bmW_j \bmF_j \Big)
            \bmJ_{j,\theta}^T  \mathbf{\Theta} \hat{\bmH}_{\theta,k}
           \\& +   \mbox{Tr}\Big(\bmK_{j,\theta}  \bmF_j^H \bmW_j \bmF_j\Big) \mbox{Tr}\Big(\bmK_{\theta,k} \mathbf{\Theta}^H \bmJ_{j,\theta}^T \mathbf{\Theta}\Big) \bmJ_{\theta,k}^T,
     \end{aligned}
    \end{equation}
    \begin{equation} \label{X_j}
    \begin{aligned}
                \bmX_j = & \Big(\hat{\bmH}_{j}^H + \hat{\bmH}_{\theta,0}^H \mathbf{\Theta}^H \hat{\bmH}_{j,\theta}^H\Big) \bmF_j^H \bmW_j \bmF_j \hat{\bmH}_j + \mbox{Tr}\Big(\bmK_j \bmF_j^H \bmW_j \bmF_j \Big) \bmJ_{j}^T
                  + \hat{\bmH}_{\theta,0}^H \mathbf{\Theta}^H \hat{\bmH}_{j,\theta}^H \bmF_j^H \bmW_j \bmF_j \hat{\bmH}_{j,\theta} \mathbf{\Theta} \hat{\bmH}_{\theta,0} \\& + \hat{\bmH}_j^H \bmF_j^H \bmW_j \bmF_j \hat{\bmH}_{j,\theta} \mathbf{\Theta} \hat{\bmH}_{\theta,0} 
                  + \mbox{Tr}\Big(\bmK_{\theta,0}  \mathbf{\Theta}^H \hat{\bmH}_{j,\theta}^H  \bmF_j^H \bmW_j \bmF_j \hat{\bmH}_{j,\theta} \mathbf{\Theta} \Big) \bmJ_{\theta,0}^T
                  + \hat{\bmH}_{\theta,0}^H \mathbf{\Theta}^H 
                   \mbox{Tr}\Big(\bmK_{j,\theta} \bmF_j^H \bmW_j \bmF_j \Big) \bmJ_{j,\theta}^T \mathbf{\Theta} \hat{\bmH}_{\theta,0}
              \\&  + \bmH_0^H \bmF_k^H \bmW_k \bmF_k \hat{\bmH}_{0,\theta} \mathbf{\Theta} \hat{\bmH}_{\theta,0}
                + \Big(\hat{\bmH}_0^H  + \hat{\bmH}_{\theta,0}^H \mathbf{\Theta}^H \hat{\bmH}_{0,\theta}^H \Big) \bmF_k^H \bmW_k \bmF_k \hat{\bmH}_0 + \mbox{Tr}\Big(\bmK_{0} \bmF_k^H \bmW_k \bmF_k\Big)  \bmJ_{0}^T  \\& + \mbox{Tr}\Big(\bmK_{j,\theta} \bmF_j^H \bmW_j \bmF_j \Big)  \mbox{Tr}\Big(\bmK_{\theta,0} \mathbf{\Theta}^H \bmJ_{j,\theta}^T  \mathbf{\Theta} \Big) \bmJ_{\theta,0}^T    
                + \hat{\bmH}_{\theta,0}^H \mathbf{\Theta}^H \hat{\bmH}_{0,\theta}^H \bmF_k^H \bmW_k \bmF_k \hat{\bmH}_{0,\theta} \mathbf{\Theta} \hat{\bmH}_{\theta,0} + \mbox{Tr}\Big(\bmK_{\theta,0} \mathbf{\Theta}^H \hat{\bmH}_{0,\theta}^H \bmF_k^H \bmW_k \bmF_k \\&\;\; \;\;\hat{\bmH}_{0,\theta}  \mathbf{\Theta} \Big) \bmJ_{\theta,0}^T
                 + \hat{\bmH}_{\theta,0}^H \mathbf{\Theta}^H \mbox{Tr}\Big(\bmK_{0,\theta} \bmF_k^H \bmW_k \bmF_k \Big)  \bmJ_{0,\theta}^T \mathbf{\Theta} \hat{\bmH}_{\theta,0}  + \mbox{Tr}\Big(\bmK_{0,\theta} \bmF_k^H \bmW_k \bmF_k\Big) \mbox{Tr}\Big(\bmK_{\theta,0} \mathbf{\Theta}^H \bmJ_{0,\theta}^T \mathbf{\Theta}\Big) \bmJ_{\theta,0}^T.
                \end{aligned}
    \end{equation} \hrulefill
    \end{subequations}
\end{figure*}
\begin{subequations}
\begin{equation} \label{UL_BF}
    \bmU_k = \Big(\bmX_k + \lambda_k \bmI\Big)^{-1} \Big(\hat{\bmH}_{k}^H + \hat{\bmH}_{\theta,k}^H \mathbf{\Theta}^H \hat{\bmH}_{0,\theta}^H\Big) \bmF_k^H \bmW_k,
\end{equation}
\begin{equation} \label{DL_BF}
    \bmV_j = \Big(\bmX_j + \lambda_0 \bmI\Big)^{-1} \Big(\hat{\bmH}_{j}^H + \hat{\bmH}_{\theta,0}^H \mathbf{\Theta}^H \hat{\bmH}_{j,\theta}^H\Big) \bmF_j^H \bmW_j,
\end{equation}
\end{subequations}
where $\bmX_k$ and $\bmX_j$ are defined in \eqref{X_k} and \eqref{X_j}, respectively, and the scalars $\lambda_k$ and $\lambda_0$ denote the Lagrange multipliers for the uplink user $k$ and the FD BS. The multipliers can be searched while performing power allocation for the users given the average total sum-power constraints. Namely, consider the singular value decomposition (SVD) of the matrices as $\bmX_{k} = \bmA_k \mathbf{\Lambda}_k \bmB_k $  and $\bmX_{j}=\bmA_j \mathbf{\Lambda}_j \bmB_j$, where $\bmA_i$ and $\bmB_i$ denote the left and right unitary matrices obtained with SVD and $\mathbf{\Lambda}_i$ denote the singular values. The average power constraints \eqref{c1} and \eqref{c2}, after some simplifications, can be written as 
\begin{subequations}
\begin{equation}
    \mbox{Tr}\Big(\bmV_j \bmV_j^H\Big) = \frac{\sum_{i=i}^{M_0} \bmS_j(i,i)} {(\lambda_0 + \mathbf{\Lambda}_j(i,i))^2},
\end{equation}
\begin{equation}
    \mbox{Tr}\Big(\bmU_k \bmU_k^H\Big) = \frac{\sum_{i=i}^{M_k} \bmS_k(i,i)} {(\lambda_j + \mathbf{\Lambda}_k(i,i))^2}.
\end{equation}
\end{subequations}
where the matrices $\bmS_k$ and $\bmS_j$ are defined as 
\begin{subequations}
\begin{equation}
\begin{aligned}
     \bmS_j= \bmB_j \Big(\hat{\bmH}_{j} + \hat{\bmH}_{j,\theta}  \mathbf{\Theta} \hat{\bmH}_{\theta,0}\Big)^H & \bmF_j^H \bmW_j \bmW_j \bmF_j \\& \Big(\hat{\bmH}_{j} + \hat{\bmH}_{j,\theta}   \mathbf{\Theta} \hat{\bmH}_{\theta,0}\Big)\bmA_j,
\end{aligned}
\end{equation}
\begin{equation}
\begin{aligned}
     \bmS_k = \bmB_k \Big(\hat{\bmH}_{k} +  \hat{\bmH}_{0,\theta}  \mathbf{\Theta} \hat{\bmH}_{\theta,k}\Big)^H & \bmF_k^H \bmW_k \bmW_k \bmF_k \\& \Big(\hat{\bmH}_{k} +  \hat{\bmH}_{0,\theta} \mathbf{\Theta} \hat{\bmH}_{\theta,k}\Big) \bmA_k.
\end{aligned}
\end{equation}
\end{subequations}
The optimal Lagrange multipliers that satisfy the average total power constraints can be determined using a linear search technique, such as the Bisection method, which we employ in this study. If the calculated values of the multipliers are found to be negative, we set them to zero. It is important to note that the average power constraint is fulfilled based on the estimated CSI.

\subsection{Passive Beamforming Under Imperfect CSI}
In this section, we delve into the optimization of the phase response of the IRS with the objective of simultaneously enhancing the UL and DL channels while mitigating the impact of SI and interference in the presence of imperfect CSI.

Let $\bmS, \bmT$, and $\bmZ$ denote the matrices independent of the IRS phase response, given in Appendix \ref{IRS-Opt-matrices}.
The expected WMMSE optimization problem for the IRS phase response $\mathbf{\Theta}$ under imperfect CSI, given the matrices $\bmS, \bmT$, and $\bmZ$, can be formally stated as

 \begin{subequations} \label{full_problem_ir} 
\begin{equation} \label{IRS_r_opt_problem}
    \underset{\substack{\mathbf{\Theta} }}{\min} \quad \mbox{Tr}\Big(\mathbf{\Theta}^H \bmZ \mathbf{\Theta} \bmT \Big) +\mbox{Tr}\Big(\mathbf{\Theta}^H \bmS^H \Big) + \mbox{Tr}\Big(\mathbf{\Theta} \bmS \Big) + c,
    \end{equation}
\begin{equation} \label{c1_r}
\text{s.t.} \quad 
  \quad   \quad  \Big|\bm{\theta}(i)\Big|=1,  \quad \forall i,
\end{equation}
\end{subequations}
where the scalar $c$ denotes the constant terms, independent of $\mathbf{\Theta}$. In \eqref{IRS_r_opt_problem}, the problem is stated with respect to the matrix $\mathbf{\Theta}$. However, we wish to maximize only the diagonal response of such matrix to maximize the ergodic WSR or minimize the expected MSE, because the off-diagonal elements result to be zero. Therefore, we first consider restating the problem \eqref{IRS_r_opt_problem} with respect to $\bm{\theta}$ instead of $\mathbf{\Theta}$. To this end, based on the result   \cite[identity 1.10.6]{zhang2017matrix}, we write the first term of \eqref{IRS_r_opt_problem} as
\begin{equation}
    \mbox{Tr}\Big(\mathbf{\Theta}^H \bmZ \mathbf{\Theta} \bmT \Big) = \bm{\theta}^H  \mathbf{\Sigma} \bm{\theta}, \quad \mbox{where}\; \mathbf{\Sigma} = \bmZ \odot \bmT^T.
\end{equation}
Let $\mathbf{s}$ denote the vector containing only the diagonal elements of the matrix $\bmS$. Then, the second and the third term of the problem \eqref{IRS_r_opt_problem} can be restated as a function of the vector $\bm{\theta}$ as 
\begin{equation}
    \mbox{Tr}\Big(\mathbf{\Theta}^H \bmS^H \Big) = {\bms}^H \bm{\theta}^*,\quad \mbox{Tr}\Big(\mathbf{\Theta} \bmS \Big) = {\bms}^T \bm{\theta}.
\end{equation}
By using the results stated above, the overall optimization problem \eqref{IRS_r_opt_problem} can be restated with respect to the vector $\bm{\theta}$ as 
\begin{subequations} \label{refprob}
\begin{equation} \label{IRS_r_opt_restated_2}
    \underset{\substack{\bm{\theta}}}{\min} \quad \bm{\theta}^H  \mathbf{\Sigma} \bm{\theta} +{\bms_r}^H \bm{\theta}^* + {\bms}^T \bm{\theta}_l ,
    \end{equation}
\begin{equation} \label{c1_rest}
\text{s.t.} \quad 
   \Big|\bm{\theta}(i)\Big| = 1,  \quad \forall i.
\end{equation}
\end{subequations}
Problem \eqref{refprob} is still very challenging as it is a non-convex problem due to the unit-modulus constraint. To solve it we adopt the majorization-maximization method \cite{pan2020multicell}. Such method aims to solve a more difficult problem by constructing a series of more tractable problems stated with the upper bound. Let $\mathcal{R}(\bm{\theta}^{(n)})$ denote the function evaluating \eqref{refprob} at the $n$-th iteration for computed $\bm{\theta}$. Let $\mathcal{R}_u(\bm{\theta}|\bm{\theta}^{(n)})$ denote an upper bound constructed at $n$-th iteration for $\mathcal{R}$. According to \cite{pan2020multicell}, for the problem of the form \eqref{refprob}, an upper bound can be constructed as

\begin{equation} \label{UB}
     \mathcal{R}_u(\bm{\theta}|\bm{\theta}^{(n)}) = 2 \mbox{Re}\{{\bms}^H \bmq^{(n)}\} + c_u,
 \end{equation}
where $c_u$  denote constant terms in the upper bound independent of $\bm{\theta}$ and $\bmq^{(n)}$ is given by 
\begin{equation} \label{qn_calcolo}
    \bmq^{(n)} = \Big(\lambda^{max} \bmI - \mathbf{  \Sigma}\Big) \bm{\theta}^{(n)} - {\bms}^*,
\end{equation}
with $\lambda^{max}$ denoting the maximum eigenvalues of $\mathbf{\Sigma}$. Based on the result above, the hard non-convex optimization problem \eqref{IRS_r_opt_restated_2} simplifies to a series of the following problem  
 \begin{subequations}  
\begin{equation} \label{IRS_r_optimized_ert}
    \underset{\substack{\bm{\theta}}}{\min} \quad  2 \mbox{Re}\{{\bms}^H \bmq^{(n)}\},
    \end{equation}
\begin{equation} \label{c1_trew}
\text{s.t.} \quad 
   \Big|\bm{\theta}(i)\Big| = 1,  \quad \forall i,
   \end{equation}
\end{subequations}
which need to be solved iteratively until convergence for each update of $\bm{\theta}$. By solving \eqref{IRS_r_optimized_ert}, we get the optimal solution of $\bm{\theta}$ at $(n+1)$-th iteration, given the $\bm{\theta}$ at the $n$-th iteration as 
\begin{equation} \label{solution_phi_r}
    \bm{\theta}^{(n+1)} = e^{i \angle\bmq^{(n)}}.
\end{equation}
When the remaining variables are held fixed during the alternating optimization process, a single update of $\bm{\theta}$ involves solving equation \eqref{solution_phi_r} until convergence. This solution provides the value of $\bm{\theta}$ that maximizes the rate. The steps for optimizing the phase response of the IRS using the majorization-maximization technique are described in detail in Algorithm \ref{IRS_opt}. The overall procedure consists of optimizing the digital beamformers, updating the weight matrices, digital combiners, and the IRS phase response through alternating optimization. The formal description of this procedure can be found in Algorithm \ref{alg_1}.

\begin{algorithm}[t]  
\caption{Robust Optimization of IRS Phase Response }\label{alg_1}
\textbf{Given:} $\bm{\theta}^0$.\\
\textbf{Initialize:}  iteration index $n=1$, accuracy $\epsilon$.\\
\textbf{Evaluate:} $\mathcal{R}(\bm{\theta}^0)$.\\
\textbf{Repeat until convergence}
\begin{algorithmic}
\STATE \hspace{0.001cm} Calculate $\bmq_i^{(n)}$ with \eqref{qn_calcolo}.
\STATE  \hspace{0.001cm} Update $\bm{\theta}_i^{(n+1)}$ with  \eqref{solution_phi_r}.\\
\STATE  \hspace{0.001cm} \textbf{if} $|\mathcal{R}^{(n+1)} - \mathcal{R}(\bm{\theta}_i^{(n)})|/\mathcal{R}(\bm{\theta}_i^{(n+1)}) \leq \epsilon$
\STATE  \hspace{0.4cm} Stop and return $\bm{\phi}_i^{(n+1)}$.
\STATE  \hspace{0.001cm} \textbf{else} n=n+1 and repeat.
\end{algorithmic}   \label{IRS_opt}
\end{algorithm}

\begin{algorithm}[t]  
\caption{Robust Beamforming for MIMO IRS-FD system}
\textbf{Given:} The CSI estimates, errors statistics and the rate weights.\\
\textbf{Initialize} the iteration index $n$, accuracy $\epsilon$, beamformers and combiners.\\
%\textbf{Evaluate:} $f(\bm{\phi}_l(0))$, \\
\textbf{Repeat until convergence}
\begin{algorithmic}
\STATE for $i$, where $i =k $ or $i=j$\\
\STATE \hspace{0.3cm} Update $\bmF_i$ with \eqref{MMSE_comb}.\\
\STATE \hspace{0.3cm} Update $\bmW_i$ with \eqref{W_matrix}.\\
\STATE \hspace{0.3cm} \textbf{if} i=k
\STATE \hspace{0.5cm} Update $\bmU_k$ with \eqref{UL_BF}.\\
\STATE \hspace{0.3cm} \textbf{else}
\STATE \hspace{0.5cm} Update $\bmV_j$ with \eqref{DL_BF}.\\
\STATE  Update $\mathbf{\Theta}$ with Algorithm  \ref{alg_1}.\\
\STATE  \textbf{if} convergence condition is satisfied\\
\STATE  \hspace{0.4cm} Stop and return the optimized variables.
\STATE  \textbf{else} repeat.
\end{algorithmic} \label{alg_1}
\end{algorithm}

\subsection{Convergence}
To prove the convergence, we consider an equivalent optimization problem for ergodic WSR maximization. For this, we consider the MSE weights and expected receive filters as new variables to be optimized, together with the digital beamformer for the DL and UL users and the IRS phase response based on the average MSE covariance matrices. The overall optimization problem as a function of the new variables can be formally stated as \cite{cirik2015weighted}
 
\begin{equation} \label{func_1}
\begin{aligned}
\underset{\substack{\bmF_k,\bmF_j,\bmU_k\\ \bmV_j,
\mathbf{\Theta},\bmW_k,\bmW_j}}{\min} &   \mbox{Tr}(\bmW_k \bmE_k) - w_k \mbox{logdet}(\frac{\mbox{ln 2}}{w_k} \bmW_k) + u_k \frac{w_k}{\mbox{ln} 2} \\& + \mbox{Tr}(\bmW_j \bmE_j) - w_j \mbox{logdet}(\frac{\mbox{ln 2}}{w_j} \bmW_j) + v_j \frac{w_j}{\mbox{ln} 2}.
\end{aligned}
\end{equation}
Consider the digital beamformers and $\mathbf{\Theta}$
to be fixed, the combiners can be chosen as the EMMSE combiners as in \eqref{MMSE_comb}.
By doing so, the new cost function has the same structure as above but with the updated average error covariance matrices, obtained by substituting the EMMSE combiners. By optimizing the new cost function with respect to the weight matrices, we can conclude that they can be optimized as

%\begin{equation}\label{func_2}
%\begin{aligned}
%\underset{\substack{\bmW_k,\bmW_j,\bmU_k\\ \bmV_j,
%\mathbf{\Theta}}}{\min} &   \mbox{Tr}(\bmW_k \bmE_k) - w_k \mbox{logdet}(\frac{\mbox{ln 2}}{w_k} \bmW_k) + u_k \frac{w_k}{\mbox{ln} 2} \\& \mbox{Tr}(\bmW_j \bmE_j) - w_j \mbox{logdet}(\frac{\mbox{ln 2}}{w_j} \bmW_j) + v_j \frac{w_j}{\mbox{ln} 2}.
%\end{aligned}
%\end{equation}

\begin{equation}
    \bmW_k = \frac{w_k}{\mbox{ln}2} \Big( \bmE_k \Big)^{-1}, \quad
        \bmW_j = \frac{w_j}{\mbox{ln}2} \Big( \bmE_j \Big)^{-1}.
\end{equation} 
\begin{figure}[t]
    \centering
\includegraphics[width=\columnwidth,height=7cm]{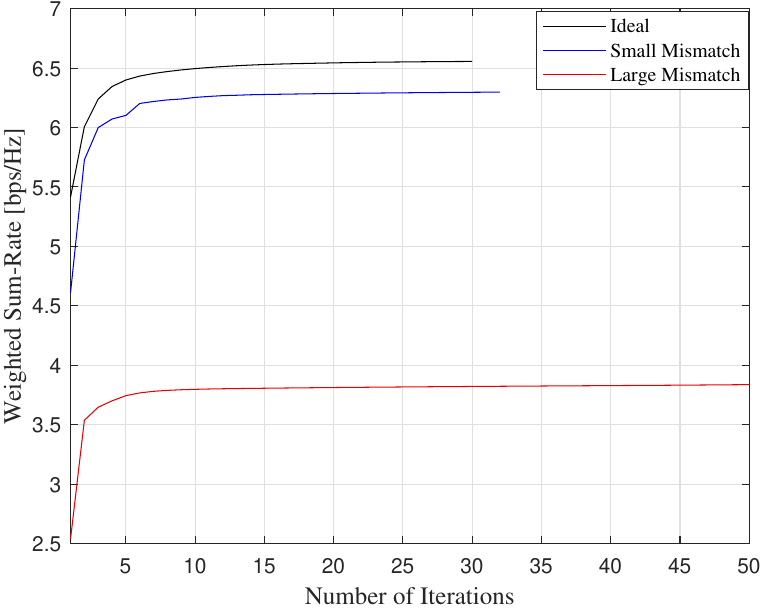}
    \caption{Convergence behaviour of the proposed design.} \label{convergenza}
\end{figure} 
By plugging the new weight matrices into the cost function above, it can be shown that the ergodic WSR maximization problem is equivalent to the original WSR cost function under the imperfect CSI, function of the average error covariance matrices. By considering the statistical distribution of the CSI errors and the CSI estimates, the proposed robust beamforming design optimizes the digital beamformers and the IRS phase response which leads to a monotonic increase in the ergodic WSR sum rate metric, which assures convergence to a local optimum \cite{joudeh2016sum,negro2012sum}.

%\begin{subequations}
%\begin{equation}
%\underset{\substack{\bmV_j,\bmU_k,\\\mathbf{\Theta}}}{\min} \quad  - w_k \mbox{ln det}\big((\bmE_k)^{-1} \big) - w_j \mbox{ln det}\big((\bmE_j)^{-1} \big)
%    \end{equation}
%\begin{equation}
%\text{s.t.}  \quad\mbox{Tr}\Big( \bmU_k \bmU_k^H \Big) 	\preceq \alpha_k,   
%\end{equation}
%\begin{equation}
%\quad \quad    \mbox{Tr} \Big( \bmV_j  \bmV_j^H \Big) \leq  %\alpha_0.  
%\end{equation}
%\begin{equation}  
% \quad 
 % \quad   \quad  \Big|\bm{\theta}(i)\Big|=1,  \quad \forall i,
%\end{equation}
 %\end{subequations}

The convergence behaviour of the robust joint active and passive beamforming design is illustrated in Fig.~\ref{convergenza}. The figure clearly demonstrates that the proposed method leads to a consistent and monotonic improvement in the WSR with each iteration, ensuring convergence. Additionally, Fig.~\ref{convergenza} depicts the convergence pattern when there is a slight or significant mismatch between the true and the estimated  CSI available. It is observed that larger uncertainties lead to less WSR and as the mismatch increases, a higher number of iterations is needed for the proposed design to converge.

%\textcolor{blue}{From the curves in Fig. 2, the convergence in all three cases looks the same.} 

\subsection{Complexity Analysis} 
In this Section, we present the computational complexity analysis for the proposed statistically robust beamforming design. Each update consists of updating the digital beamformers for the DL and UL users, searching for the Lagrange multipliers and finally updating the IRS phase response based on solving sub-problems.
 
The complexity of updating the digital beamformers for the DL and UL users can be expressed as $\mathcal{O}(N_j^3)$ and $\mathcal{O}(N_0^3)$, respectively. The complexity of the Lagrange multipliers search is considered negligible since it is linear. In order to update the phase response of the IRS, the majorization-minimization approach requires the initial calculation of the maximum eigenvalue for the matrix $\mathbf{\Sigma}$, which has a complexity of $(RC)^3$. During each update of the phase response, the main computational burden lies in computing $\bmq$ at each iteration, which has a complexity of $(RC)^2$. Let $N_{max}$ denote the total number of iterations required for each IRS update when updating the digital beamformers. The overall complexity can be expressed as $\mathcal{O}(N_j^3 + N_0^3 + (RC)^3 + N_{max} (RC)^2)$.

\section{Numerical Results} \label{risultati}
In this section, we present extensive simulation results to evaluate the performance of the proposed robust beamforming design. The FD BS and the IRS are positioned at $(0\text{m},0\text{m},0\text{m})$ and $(20\text{m},10\text{m},0\text{m})$, respectively, in three-dimensional coordinates. The UL and DL users are randomly distributed within circular regions of radius $r=8$~m centered at $(20\text{m},0\text{m},30\text{m})$ and $(30\text{m},0\text{m},20\text{m})$, respectively. Both the FD BS and the users are equipped with uniform linear arrays (ULAs) positioned at a half-wavelength apart. The IRS consists of $10 \times 10 = 100$ elements and assists in MIMO FD communication unless stated otherwise. The MIMO FD BS has $M_0=15$ transmit and $N_0=8$ receive antennas. The UL user $k$ and the DL user $j$ are assumed to have $M_k=5$ transmit and $N_j=5$ receive antennas, respectively, and each user is served with $d_k=d_j=2$ data streams. For simulations, the digital beamformers are initialized as the dominant eigenvectors of the effective channel covariance matrix of the intended user and the IRS phase response is initialized to be random. 
%\begin{figure}
%    \centering%
%\includegraphics[width=1\columnwidth,height=5cm]{scenario_simulazione.pdf}
 %   \caption{Simulation scenario to test the performance of the proposed robust beamforming design.}
  % \label{fig:_IRS_conv}
%\end{figure} 

In general, the wireless channel models can be obtained using stochastic or deterministic methods, or by using their combination. Deterministic modelling usually employs ray tracing-based methods, as in \cite{he2018design,guan20205g}. These deterministic channel models can achieve high accuracy, yet require substantial data for the characterization of real-world surroundings. Therefore, we adopt stochastic modelling for channel characterization, which is also widely adopted in the literature. 
More specifically, to model the large-scale fading, we adopt the following path-loss model \cite{pan2020multicell}
\begin{equation}
    PL = PL_0 - 10 \alpha \log_{10}\left(\frac{d}{d_0}\right),
\end{equation}
where $PL_0=-30~$dB is the pathloss at the reference distance $d_0 = 1~$m, and $\alpha$ is the path-loss exponent set to be $2$. The SI channel is modelled according to the Rician fading channel model \cite{duarte2012experiment}

\begin{equation}
    \bmH_0 = \sqrt{\frac{\kappa_0}{\kappa_0 + \kappa_0}} \bmH_0^{LoS} + \sqrt{\frac{1}{\kappa_0 + 1}} \bmH_0^{NLoS} 
\end{equation}
where $\kappa_0 = 1~$ denotes the Rician factor, $\bmH_0^{LoS}$ is the deterministic line-of-sight (LoS) component and $\bmH_0^{Ref} $ denote the non-LoS (NLoS) component which is Rayleigh fading. The direct links between the users and the FD BS are also modelled with a Rician fading channel model with a Rician factor of $1$. The channels involving the IRS are modelled according to Rayleigh fading \cite{guo2019weighted}.

We define the signal-to-noise (SNR) of our system as 

\begin{equation}
    SNR = \frac{\alpha_0}{\sigma_j^2} = \frac{\alpha_k}{\sigma_0^2},
\end{equation}
where $\alpha_0$ and $\alpha_k$ are the average total transmit power at the FD BS and the multi-antenna UL user $k$, respectively. We assume that the CSI errors to be i.i.d zero-mean  circularly
symmetric complex Gaussian distribution with the same variance $\sigma_{csi}^2$, and therefore, the error covariance matrices set chosen as 

\begin{equation} \label{Channel_modelling}
    \begin{cases}
        \bmJ_k = \bmI, \;\bmK_k = \sigma_{csi}^2 \bmI, \quad \quad  \quad \quad  \bmJ_{\theta,0} = \bmI,\; \bmK_{\theta,0} =  \sigma_{csi}^2 \bmI  ,  \\
     \; \bmJ_0 = \bmI, \;\bmK_0 = \sigma_{csi}^2 \bmI, \quad  \quad \quad  \quad \bmJ_{j,k} = \bmI,  \bmK_{j,k}= \sigma_{csi}^2 \bmI, \\ \;\bmJ_j  = \bmI, \bmK_j= \sigma_{csi}^2 \bmI, 
  \quad \quad \quad  \quad \; \bmJ_{0,\theta} = \bmI,  \bmK_{0,\theta}= \sigma_{csi}^2 \bmI \\
 \; \bmJ_{j,\theta} = \bmI,   \bmK_{j,\theta}=\sigma_{csi}^2 \bmI  \quad  \quad \quad \bmJ_{\theta,k}= \bmI,   \bmK_{\theta,k}=\sigma_{csi}^2 \bmI\\
    \end{cases}       
\end{equation}

\begin{figure}[t]
     \centering
\includegraphics[width=\columnwidth,height=7cm]{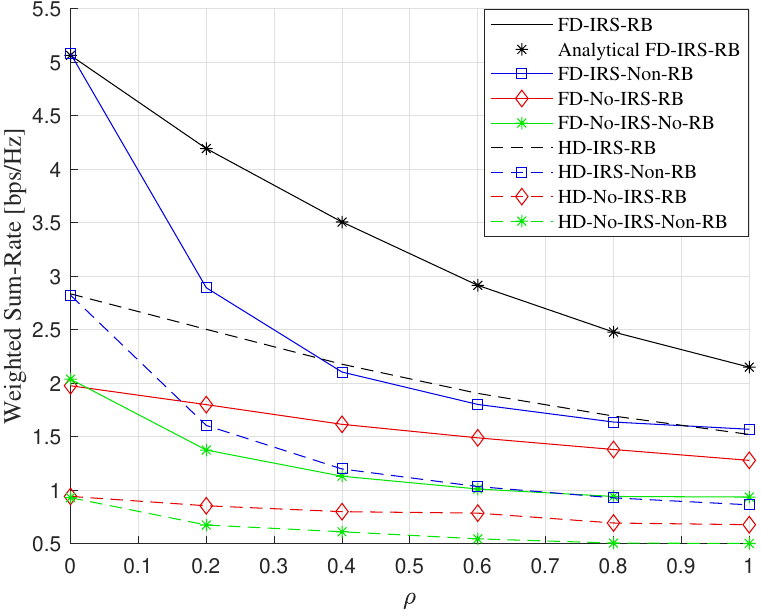}
    \caption{Average WSR as a function of $\rho$  with SNR$=30~$dB.} \label{WSR_roh}  \vspace{-3mm}
\end{figure}

Since the variance of the CSI errors strictly depend on the average transmit power (used also to estimate the channels)  and the noise variance, we assume that $\sigma_{csi}^2$ decays as $\mathcal{O}(SNR^{-\alpha})$, for some constant $\alpha$ \cite{joudeh2016sum}, satisfying the CSI error decay rate inversely proportional to the SNR. As the SNR represents the transmit SNR, note that $SNR \rightarrow \infty$ is equivalent to $\alpha_k=\alpha_0 \rightarrow \infty$, which enhance the CSI quality and reduces the CSI error variance as $\sigma_{csi}^2 \rightarrow 0$.  The CSI errors variance is set as $\sigma_{csi}^2 = \rho /SNR^\alpha$, where $\rho$ denotes a scale factor and $\alpha \in [0,1]$ determine the quality of the CSI. Maintaining the quality of the CSI with $\alpha=1$ could be exhausting in terms of resources required to acquire the CSI and therefore we set $\alpha = 0.6$ \cite{joudeh2016sum}.

We label our proposed design as a \emph{FD-IRS-RB}.
For comparison, we define the following benchmark schemes:
\begin{enumerate}
    \item \emph{FD-IRS-Non-RB}: FD system assisted with IRS and non robust beamforming, i.e., the available CSI is treated as perfect;
    \item \emph{FD-No-IRS-RB}: FD system with no IRS and robust beamforming;
    \item \emph{FD-No-IRS-Non-RB}: FD system with no IRS and non robust beamforming;
    \item \emph{HD-IRS-RB}: HD system assisted with IRS and robust beamforming;
    \item \emph{HD-IRS-Non-RB}: HD system assisted with IRS and non robust beamforming;
    \item \emph{HD-No-IRS-RB}: HD system with no IRS and robust beamforming;
    \item \emph{HD-No-IRS-Non-RB}: HD system with no IRS and non robust beamforming.
\end{enumerate}
We also compare our approximation proposed for the ergodic WSR in Theorem \ref{theorem_1} given the transmit covariance matrices of the proposed robust beamforming design which we label as \emph{Analytical FD-IRS-RB} and compare it with the ergodic rate achieved with the EWMMSE.

\begin{figure*}[t]
    \centering
 \begin{minipage}{0.49\textwidth}
      \centering
\includegraphics[width=\columnwidth,height=7cm]{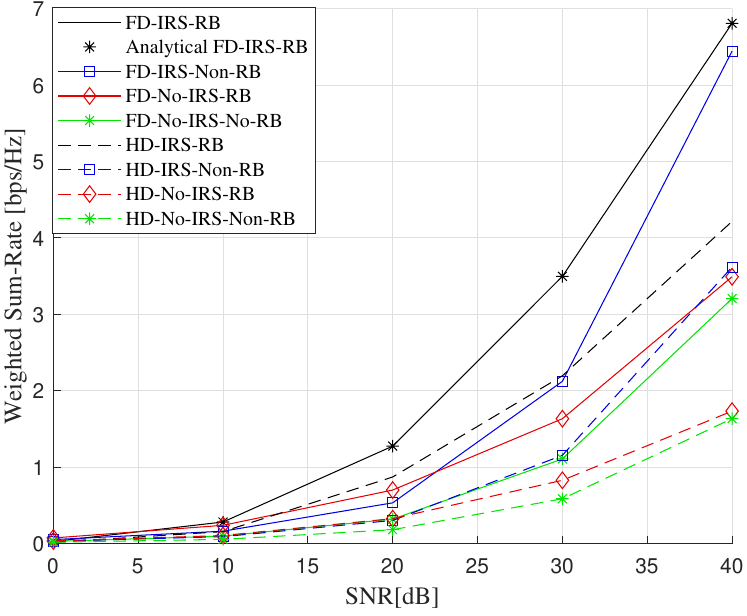}
    \caption{Average WSR as a function of SNR with $\rho=0.4$.}   \label{WSR_roh_04}
    \end{minipage} \hfill
      \begin{minipage}{0.49\textwidth}
     \centering
\includegraphics[width=\columnwidth,height=7cm]{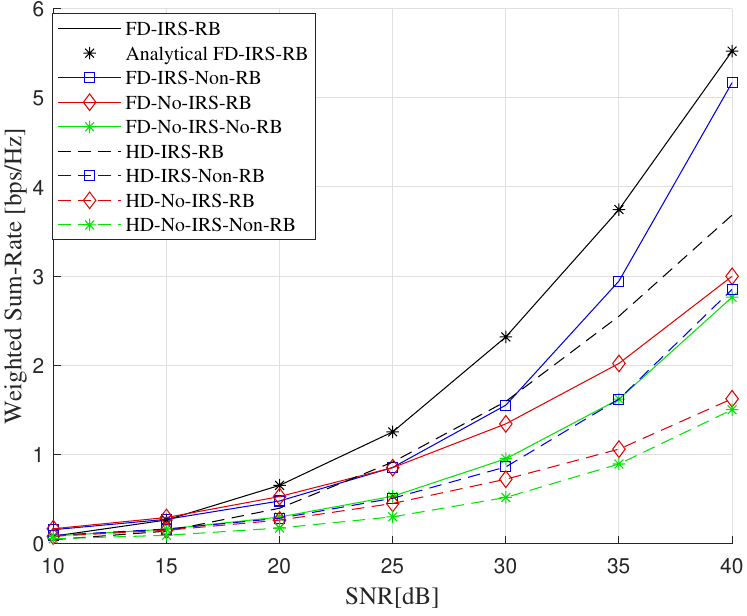}
    \caption{Average WSR as a function of SNR with $\rho=0.9$.} 
    \label{WSR_roh_09}
    \end{minipage}\hfill    \vspace{-3mm}
\end{figure*}

\begin{figure*}[t]
    \centering
 \begin{minipage}{0.49\textwidth}
 \centering
\includegraphics[width=\columnwidth,height=7cm]{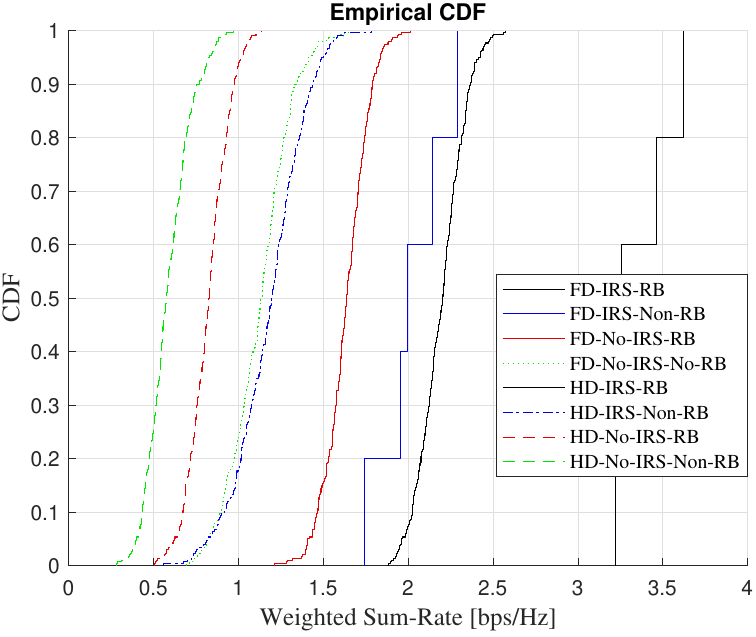}
    \caption{CDF of the various schemes with $\rho=0.4$.} 
    \label{CDF_04}
    \end{minipage} \hfill
      \begin{minipage}{0.49\textwidth}
     \centering
\includegraphics[width=\columnwidth,height=7cm]{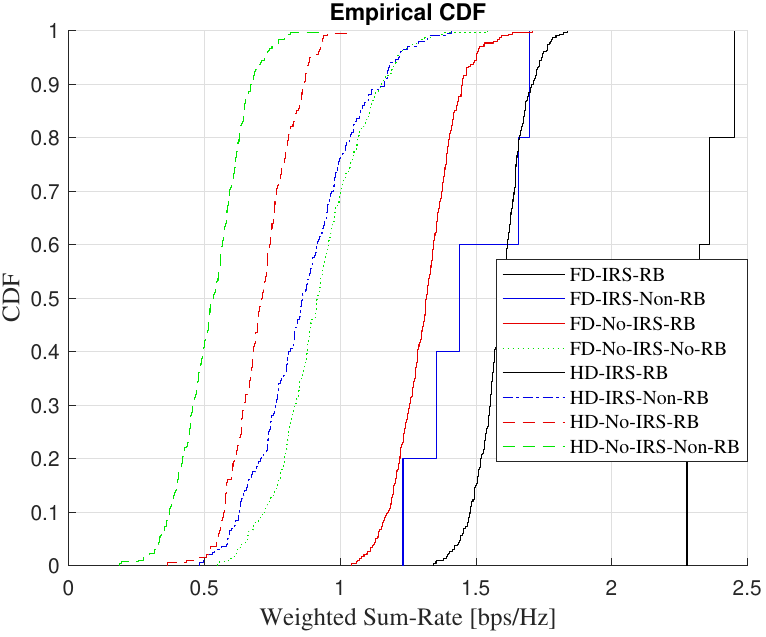}
    \caption{CDF of the various schemes with $\rho=0.9$.}
      \label{CDF_09}   
    \end{minipage}\hfill  \vspace{-3mm}
\end{figure*}

Fig. \ref{WSR_roh} shows the performance of the proposed robust joint beamforming design as a function of the scale factor $\rho$ dictating the CSI error variance, by means of Monte Carlo simulations at SNR$=30~$dB. It is shown that the proposed robust design achieves significant performance gain compared to the naive FD-IRS-Non-RB scheme, which does not account for the CSI errors. Moreover, we can also see that the achieved ergodic WSR accurately matches the result stated in Theorem \ref{theorem_1}. It is to be noted that as the CSI error variance gets extremely large, the proposed statistically robust beamforming design preserves significant robustness against the uncertainties in the CSI quality.
Namely, for $\rho \geq 0.4$, the FD-IRS-Non-RB scheme achieves less gain than the IRS-aided HD system deploying robust beamforming. Such a result showcases the importance of adopting a robust beamforming approach for the IRS-FD systems, as, in practice, their performance could degrade significantly due to imperfect CSI.

% Fig.~\ref{WSR_roh_04} and Fig.~\ref{WSR_roh_09} show the ergodic WSR of the proposed design as a function of the transmit SNR compared with the benchmark schemes with $\rho =0.4$ and $\rho=0.9$, respectively. We can see that out design achieves significant gain in the presence of CSI errors compared to the other schemes. Theoretically, the FD systems offer a twofold gain in the WSR compared to the HD systems. However, Figures \ref{WSR_roh_04}-\ref{WSR_roh_09} for the scheme FD-IRS-Non-RB, at low SNR, i.e. high CSI error variance, show that the IRS-FD system may achieve no gain compared to the IRS-aided HD system  at any SNR level if non-robust beamforming approach is adopted in the case of uncertainties in the CSI. This is because the residual SI and the cross-interference are crucial in determining their gain in the UL and DL directions, and with large CSI error variance, they cannot be handled adequately. Moreover, in the presence of significant CSI errors, the gain of the FD systems is very much limited and far from the theoretical two-fold expected gain. However, we can observe that as the CSI error variance decreases, the performance of the non-robust beamforming scheme tends toward the performance of the robust beamforming schemes, which tends towards the two-fold improvement in the WSR due to quasi-ideal simultaneous transmission and reception. 

Fig.~\ref{WSR_roh_04} and Fig.~\ref{WSR_roh_09} present the achieved ergodic WSR of our proposed design as a function of the transmit SNR, compared to benchmark schemes with $\rho=0.4$ and $\rho=0.9$, respectively. The results clearly demonstrate that our design achieves significant gains in the presence of CSI errors when compared to the other schemes. Theoretically, FD systems offer a twofold improvement in the WSR compared to HD systems. However, the corresponding curves for the FD-IRS-Non-RB scheme reveal that with high CSI error variance, the IRS-FD system may exhibit no gain in comparison to the IRS-aided HD system at any SNR level if a non-robust beamforming approach is employed in the presence of large CSI uncertainties. This can be attributed to the crucial role played by residual SI in determining the IRS-FD system's gain. With a large CSI error variance, these interference components cannot be adequately handled. On the other hand, adopting robust beamforming for IRS-FD systems can be promising in terms of robustness. We can also observe that as the CSI error variance decreases, the performance of the non-robust beamforming scheme gradually converges towards that of the robust beamforming schemes, approaching the desired twofold improvement in the WSR due to quasi-ideal simultaneous transmission and reception.

Fig.~\ref{CDF_04} and Fig.~\ref{CDF_09} show the cumulative distribution function (CDF) of the proposed statistically robust beamforming design, in comparison with the benchmark schemes at $SNR = 30~$dB. It is clearly visible that the proposed method achieves significant average WSR compared to the other schemes and the reported gains are stable with different values of the scale factor $\rho$, and therefore with the different amount of CSI error variance. 

%The non-robust IRS-FD system shows similar average performance as the HD system assisted with IRS with robust beamforming, which motivates in adopting a robust beamforming approach to leverage the full potential of the IRS-FD systems and pave the path towards practical highly spectrally and energy-efficient communication systems.

Based on the presented numerical results, it can be concluded that CSI errors play a critical role in determining the performance of IRS-FD systems. Neglecting these errors can significantly impact the system's performance, leading to suboptimal beamforming decisions. Consequently, the potential of the IRS-FD can not be fully exploited, limiting its ability to enhance channel quality for UL and DL users and mitigate interference. In the UL direction, uncertainties in the SI channel result in residual SI that cannot be adequately suppressed even with joint active and passive beamforming, resulting in a low signal-to-total-interference plus noise ratio. In the DL direction, FD systems experience interference from nearby UL users, which can be as detrimental as well. Therefore, effective mitigation of interference through robust beamforming decisions is crucial for achieving satisfactory UL and DL performance. The results demonstrate that adopting a statistically robust beamforming approach is essential for effectively managing SI and cross-interference in practical IRS-FD systems with imperfect CSI. By doing so, the full potential of IRS-FD systems can be harnessed, leading to highly efficient communication systems in terms of spectral and energy efficiency.
  
\section{Conclusions} \label{Conc}
In this paper, the authors introduced a novel statistically robust joint beamforming design for maximizing the ergodic WSR of an IRS-FD system in the presence of imperfect CSI. The CSI errors were modelled using the Gaussian-Kronecker model, and an approximation for the ergodic WSR was derived based on the statistical distribution of the errors given the channel estimates. To address the ergodic WSR maximization problem, an iterative approach based on the EWMMSE method was employed, which involved solving two layers of sub-problems through alternating optimization. Simulation results demonstrated that imperfect CSI could significantly impact the performance of IRS-FD systems, and adopting a robust beamforming strategy led to substantial improvements compared to the naive approach. Furthermore, the derived approximation for the ergodic WSR aligned well with the achieved results using the expected WMMSE method.
 
 \vspace{-3mm}
\appendices 
\section{Auxilary Matrices for \eqref{aux_Q}} \label{aux-matrices}
The matrices $\bmQ_k, \bmT_j, \bmT_{0}, \bmQ_{j,k}$ are defined as follows
\begin{equation}
\begin{aligned}
     \bmQ_k & =   \mathbb{E}_{\bmH|\hat{\bmH}}\Big(\overline{\bmH}_k \widetilde{\bmU}_k \overline{\bmH}_k^H\Big) \\&
      =   \hat{\bmH}_k \widetilde{\bmU}_k \hat{\bmH}_k^H + \hat{\bmH}_k \widetilde{\bmU}_k \hat{\bmH}_{\theta,k}^H \mathbf{\Theta}^H \hat{\bmH}_{0,\theta}^H + 
      \mbox{Tr}\Big(\widetilde{\bmU}_k \bmJ_k^T\Big) \bmK_k 
      \\& \;\; + \hat{\bmH}_{0,\theta} \mathbf{\Theta}  \hat{\bmH}_{\theta,k} \widetilde{\bmU}_k \hat{\bmH}_{\theta,k}^H \mathbf{\Theta}^H \hat{\bmH}_{0,\theta}^H + \hat{\bmH}_{0,\theta} \mathbf{\Theta} \hat{\bmH}_{\theta,k} \widetilde{\bmU}_k \hat{\bmH}_{k}^H \\&\;\;
     + \hat{\bmH}_{0,\theta} \mathbf{\Theta} \mbox{Tr}\Big(\widetilde{\bmU}_k \bmJ_{\theta,k}^T\Big) \bmK_{\theta,k} \mathbf{\Theta}^H \hat{\bmH}_{0,\theta}^H
  \\& \;\;  + \mbox{Tr}\Big(\mathbf{\Theta} \hat{\bmH}_{\theta,k} \widetilde{\bmU}_k \hat{\bmH}_{\theta,k}^H \mathbf{\Theta}^H \bmJ_{0,\theta}\Big) \bmK_{0,\theta}
  \\&
  \;\; + \mbox{Tr}\Big(\widetilde{\bmU}_k \bmJ_{\theta,k}^T\Big) \mbox{Tr}\Big(\mathbf{\Theta}  \bmK_{\theta,k}  \mathbf{\Theta}^H \bmJ_{0,\theta}^T\Big) \bmK_{0,\theta},
\end{aligned}
\end{equation}
\begin{equation}
\begin{aligned}
     \bmT_j & =  \mathbb{E}_{\bmH|\hat{\bmH}}\Big(\overline{\bmH}_j \widetilde{\bmV}_j \overline{\bmH}_j^H\Big) \\&
      =  \hat{\bmH}_j \widetilde{\bmV}_j \hat{\bmH}_j^H + \hat{\bmH}_j \widetilde{\bmV}_j  \hat{\bmH}_{\theta,0}^H \mathbf{\Theta}^H \hat{\bmH}_{j,\theta}^H + \mbox{Tr}\Big(\widetilde{\bmV}_j \bmJ_j^T\Big) \bmK_j
      \\& \; \;+ \hat{\bmH}_{j,\theta} \mathbf{\Theta} \hat{\bmH}_{\theta,0} \widetilde{\bmV}_j \hat{\bmH}_j^H + \hat{\bmH}_{j,\theta} \mathbf{\Theta} \hat{\bmH}_{\theta,0} \widetilde{\bmV}_j \hat{\bmH}_{\theta,0}^H \mathbf{\Theta}^H \hat{\bmH}_{j,\theta}^H \\& \;\; \;+ \hat{\bmH}_{j,\theta} \mathbf{\Theta} \mbox{Tr}\Big(\widetilde{\bmV}_j \bmJ_{\theta,0}^T\Big) \bmK_{\theta,0} \mathbf{\Theta}^H \hat{\bmH}_{j,\theta}^H \\&  \;\;\;+ 
      \mbox{Tr}\Big( \mathbf{\Theta} \hat{\bmH}_{\theta,0} \widetilde{\bmV}_j \hat{\bmH}_{\theta,0}^H \mathbf{\Theta}^H \bmJ_{j,\theta}^T \Big) \bmK_{j,\theta}\\&  \;\;\;+ 
     \mbox{Tr}\Big(\widetilde{\bmV}_j \bmJ_{\theta,0}^T\Big) \mbox{Tr}\Big(\mathbf{\Theta} 
       \bmK_{\theta,0}
      \mathbf{\Theta}^H  \bmJ_{j,\theta}^T \Big) \bmK_{j,\theta},
      %\Delta \bmH_{j,\theta} \mathbf{\Theta} \Delta \hat{\bmH}_{\theta,0} \widetilde{\bmV}_j \Delta \hat{\bmH}_{\theta,0}^H \mathbf{\Theta}^H \Delta \bmH_{j,\theta}^H
\end{aligned}
\end{equation}

\begin{equation}
\begin{aligned}
        \bmT_{0}& = \mathbb{E}_{\bmH|\hat{\bmH}}\Big(\overline{\bmH}_0 \widetilde{\bmV}_j \overline{\bmH}_0^H\Big)  \\&
      =  \hat{\bmH}_0 \widetilde{\bmV}_j \hat{\bmH}_0^H  + \hat{\bmH}_0 \widetilde{\bmV}_j \hat{\bmH}_{\theta,0}^H \mathbf{\Theta}^H \hat{\bmH}_{0,\theta}^H + \mbox{Tr}\Big(\widetilde{\bmV}_j \bmJ_0^T\Big) \bmK_0^T
      \\&  \;\;
    + \hat{\bmH}_{0,\theta} \mathbf{\Theta} \hat{\bmH}_{\theta,0} \widetilde{\bmV}_j \hat{\bmH}_{0}^H + \hat{\bmH}_{0,\theta} \mathbf{\Theta} \hat{\bmH}_{\theta,0} \widetilde{\bmV}_j \hat{\bmH}_{\theta,0}^H \mathbf{\Theta}^H \hat{\bmH}_{0,\theta}^H \\&  \;\;
    + \hat{\bmH}_{0,\theta} \mathbf{\Theta} \mbox{Tr}\Big(\widetilde{\bmV}_j \bmJ_{\theta,0}^T\Big) \bmK_{\theta,0}
     \mathbf{\Theta}^H \hat{\bmH}_{0,\theta}^H \\&  \;\;
    + \mbox{Tr}\Big(\mathbf{\Theta}  \hat{\bmH}_{\theta,0} \widetilde{\bmV}_j  \hat{\bmH}_{\theta,0}^H \mathbf{\Theta}^H \bmJ_{0,\theta}^T\Big) \bmK_{0,\theta}
    \\&  \;\;
    +  \mbox{Tr}(\widetilde{\bmV}_j \bmJ_{\theta,0}^T)  \mbox{Tr}\Big(\mathbf{\Theta} \bmK_{\theta,0} \mathbf{\Theta}^H \bmJ_{0,\theta}^T\Big) \bmK_{0,\theta},
\end{aligned}
\end{equation}

\begin{equation}
\begin{aligned}
   \bmQ_{j,k} & =  \mathbb{E}_{\bmH|\hat{\bmH}}\Big(\overline{\bmH}_{j,l}  \widetilde{\bmU}_l \overline{\bmH}_{j,k}^H\Big) \\&   = \hat{\bmH}_{j,k} \widetilde{\bmU}_k \hat{\bmH}_{j,k}^H + \hat{\bmH}_{j,k} \widetilde{\bmU}_k \hat{\bmH}_{\theta,k}^H \mathbf{\Theta}^H \hat{\bmH}_{j,\theta}^H + \mbox{Tr}\Big(\widetilde{\bmU}_k \bmJ_{j,k}^T\Big) \bmK_{j,k}
   \\& \;\;+
   \hat{\bmH}_{j,\theta} \mathbf{\Theta} \hat{\bmH}_{\theta,k} \widetilde{\bmU}_k \hat{\bmH}_{j,k}^H +  \hat{\bmH}_{j,\theta} \mathbf{\Theta} \hat{\bmH}_{\theta,k} \widetilde{\bmU}_k  \hat{\bmH}_{\theta,k}^H \mathbf{\Theta}^H \hat{\bmH}_{j,\theta}^H \\& \;\;+ 
   + \hat{\bmH}_{j,\theta} \mathbf{\Theta} 
   \mbox{Tr}\Big(\widetilde{\bmU}_k \bmJ_{\theta,k}^T\Big) \bmK_{\theta,k}
    \mathbf{\Theta}^H \hat{\bmH}_{j,\theta}^H \\& \;\;+ 
    \mbox{Tr}\Big(\mathbf{\Theta} \hat{\bmH}_{\theta,k} \widetilde{\bmU}_k \hat{\bmH}_{\theta,k}^H \mathbf{\Theta}^H \bmJ_{j,\theta}^T \Big) \bmK_{j,\theta} \\& \;\;+ 
    \mbox{Tr}\Big(\widetilde{\bmU}_k \bmJ_{\theta,k}^T\Big) \mbox{Tr}\Big(\mathbf{\Theta}   \bmK_{\theta,k} \mathbf{\Theta}^H \bmJ_{j,\theta}^T \Big) \bmK_{j,\theta}.
   \end{aligned}
\end{equation}

 \section{Matrices to optimize $\mathbf{\Theta}$} \label{IRS-Opt-matrices}
The matrices $\bmS, \bmT$, and $\bmZ$ to optimize the IRS phase response $\mathbf{\Theta}$ are as follows:
 
\begin{equation}
 \begin{aligned}
     \bmS =& \hat{\bmH}_{0,\theta}^H \bmF_k^H \Big(\bmW_k \bmF_k \hat{\bmH}_k \widetilde{\bmU}_k \hat{\bmH}_{\theta,k}^H +  \bmW_k \bmF_k \hat{\bmH}_0 \widetilde{\bmV}_j \hat{\bmH}_{\theta,0}^H  \\& + \bmW_k \bmU_k^H \hat{\bmH}_{\theta,k}^H \Big) +  \hat{\bmH}_{j,\theta}^H \bmF_j^H \Big(\bmW_j \bmF_j \bmH_j \widetilde{\bmV}_j \bmH_{\theta,0}^H  \\& + \bmW_j \bmF_j \bmH_{j,k} \widetilde{\bmU}_k \bmH_{\theta,k}^H - \bmW_j \bmV_j^H \hat{\bmH}_{\theta,0}^H \Big),
 \end{aligned}
 \end{equation}

 \begin{equation}
 \begin{aligned}
       \bmZ = & \hat{\bmH}_{0,\theta}^H \bmF_k^H \bmW_k \bmF_k \bmH_{0,\theta} + \hat{\bmH}_{0,\theta}^H \bmF_k^H \bmW_k \bmF_k  \bmH_{0,\theta} \\& + 
       \mbox{Tr}\Big(\bmF_k^H \bmW_k \bmF_k \bmJ_{0,\theta}^T\Big) \bmK_{0,\theta} +  \mbox{Tr}\Big( \bmF_k^H \bmW_k \bmF_k \bmJ_{0,\theta}^T\Big) \bmK_{0,\theta} \\& + \mbox{Tr}\Big(\bmF_k^H \bmW_k \bmF_k \bmJ_{0,\theta}^T\Big) \bmK_{0,\theta} + \hat{\bmH}_{0,\theta}^H \bmF_k^H \bmW_k \bmF_k \hat{\bmH}_{0,\theta} \\& + \hat{\bmH}_{0,\theta}^H \bmF_k^H \bmW_k \bmF_k \hat{\bmH}_{0,\theta} + 
       \mbox{Tr}\Big( \bmF_k^H \bmW_k \bmF_k \bmJ_{0,\theta}^T\Big) \bmK_{0,\theta} \\& +\hat{\bmH}_{j,\theta}^H \bmF_j \bmW_j \bmF_j \hat{\bmH}_{j,\theta} + \hat{\bmH}_{j,\theta}^H \bmF_j^H \bmW_j \bmF_j \hat{\bmH}_{j,\theta} \\& + 
       \mbox{Tr}\Big(\bmF_j^H \bmW_j \bmF_j \bmJ_{j,\theta}^T\Big) \bmK_{j,\theta}
        + \mbox{Tr}\Big(\bmF_j^H \bmW_j \bmF_j \bmJ_{j,\theta}^T\Big) \bmK_{j,\theta} \\& +
         \hat{\bmH}_{j,\theta}^H \bmF_j^H \bmW_j \bmF_j \hat{\bmH}_{j,\theta} +  \hat{\bmH}_{j,\theta}^H \bmF_j^H \bmW_j \bmF_j \hat{\bmH}_{j,\theta} \\& +
         \mbox{Tr}\Big(\bmF_j^H \bmW_j \bmF_j \bmJ_{j,\theta}^T\Big) \bmK_{j,\theta} +  
         \mbox{Tr}\Big(\bmF_j^H \bmW_j \bmF_j \bmJ_{j,\theta}^T\Big) \bmK_{j,\theta}, 
 \end{aligned}
 \end{equation}

\begin{equation}
\begin{aligned}
    \bmT = & \hat{\bmH}_{\theta,k} \widetilde{\bmU}_k \hat{\bmH}_{\theta,k}^H + \mbox{Tr}\Big(\widetilde{\bmU}_k \bmJ_{\theta,k}^T\Big) \bmK_{\theta,k} + \hat{\bmH}_{\theta,k} \widetilde{\bmU}_k \hat{\bmH}_{\theta,k}^H \\& + \mbox{Tr}\Big(\widetilde{\bmU}_k \bmJ_{\theta,k}^T\Big) \bmK_{\theta,k} + \hat{\bmH}_{\theta,0} \widetilde{\bmV}_j \hat{\bmH}_{\theta,0}^H + 
  \mbox{Tr}\Big(\widetilde{\bmV}_j \bmJ_{\theta,0}^T\Big) \bmK_{\theta,0}
    \\& + \hat{\bmH}_{\theta,0} \widetilde{\bmV}_j  \hat{\bmH}_{\theta,0}^H  + \mbox{Tr}\Big(\widetilde{\bmV}_j \bmJ_{\theta,0}^T\Big) \bmK_{\theta,0} +   \hat{\bmH}_{\theta,0} \widetilde{\bmV}_j  \hat{\bmH}_{\theta,0}^H \\& + \mbox{Tr}\Big(\widetilde{\bmV}_j \bmJ_{\theta,0}^T\Big) \bmK_{\theta,0} + \hat{\bmH}_{\theta,k} \widetilde{\bmU}_k \hat{\bmH}_{\theta,k}^H + \mbox{Tr}\Big(\widetilde{\bmU}_k \bmJ_{\theta,k}^T\Big) \bmK_{\theta,k} \\& + \hat{\bmH}_{\theta,k} \widetilde{\bmU}_k \hat{\bmH}_{\theta,k}^H + \mbox{Tr}\Big(\widetilde{\bmU}_k \bmJ_{\theta,k}^T\Big) \bmK_{\theta,k}.
\end{aligned}
\end{equation}

 \section*{Acknowledgement}
This work is supported by the Luxembourg National Fund (FNR)-RISOTTI–the Reconfigurable Intelligent Surfaces for Smart Cities under Project FNR/C20/IS/14773976/RISOTTI.

\def\baselinestretch{0.97}
\bibliographystyle{IEEEtran}
\bibliography{main}

\begin{IEEEbiography}[{\includegraphics[width=1in,height=1.25in,clip]{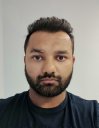}}]{\textbf{Chandan Kumar Sheemar}}(Member, IEEE) received his PhD from EURECOM, Sophia Antipolis, France, in 2022. He is currently a research associate at the Interdisciplinary Centre for Security, Reliability and Trust (SnT), University of Luxembourg, Luxembourg., within the SIGCOM research group. His main interests include full duplex, joint communications and sensing, massive MIMO, reflecting intelligent surfaces, optimization theory, and non-terrestrial networks.
\end{IEEEbiography}
 
\begin{IEEEbiography}[{\includegraphics[width=1in,height=1.25in,clip]{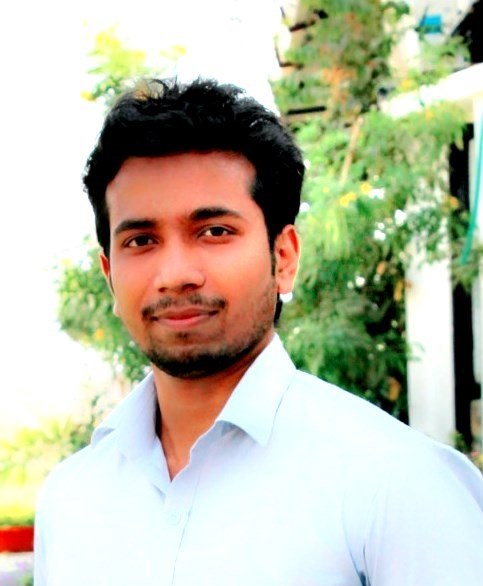}}]{\textbf{Sourabh Solanki}}(Member, IEEE) received the M. Tech. degree in Communication and Signal Processing, and a PhD degree in Electrical Engineering, both from the Indian Institute of Technology (IIT) Indore, India, in 2015 and 2019, respectively.
From 2019 to 2020, he was a Research Professor at the Korea University, Seoul, South Korea, where he received the Brain Korea 21 Postdoctoral Fellowship from the National Research Foundation, Government of Korea. Since 2021, he has been working as a
Research Associate with the Interdisciplinary Centre for Security, Reliability and Trust (SnT), University of Luxembourg, Luxembourg. His main research interests include cognitive radio, terahertz communications, mmWave networks, energy harvesting, and non-terrestrial networks. He has been serving as a technical program committee member of various conferences and has also been involved in the peer review process of major IEEE journals and conferences. He was a co-recipient of the Best Paper Award at the International Conference on ICT Convergence (ICTC), Jeju Island, South Korea, in October 2020. He also received an exemplary reviewer award for IEEE Transactions on Communications in 2020.
\end{IEEEbiography}

\begin{IEEEbiography}[{\includegraphics[width=1in,height=1.25in,clip]{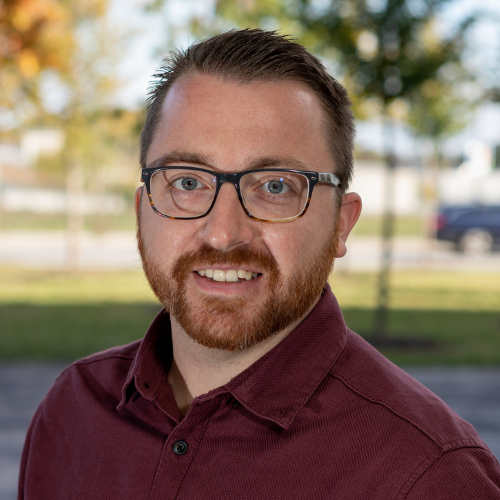}}]{\textbf{Jorge Querol}}(Member, IEEE) received the
B.Sc. (+5) degree in telecommunication engineering, the M.Sc. degree in electronics engineering, the M.Sc. degree in photonics, and the PhD degree (cum laude) in signal processing and communications from Universitat Politècnica de Catalunya-BarcelonaTech (UPC), Barcelona, Spain, in 2011, 2012, 2013, and 2018, respectively. Since 2018, he has been with the SIGCOM Research
Group, Interdisciplinary Centre for Security, Reliability, and Trust (SnT), University of Luxembourg, Luxembourg, and the Head of the Satellite Communications Laboratory. He is involved in several ESA and Luxembourgish national research projects dealing with signal processing and satellite communications. His research interests include SDR, real-time signal processing, satellite communications, 5G non-terrestrial networks, satellite navigation, and remote sensing. He received the Best Academic Record Award of the year in electronics engineering from UPC, in 2012, the First Prize of the European Satellite Navigation Competition (ESNC)
Barcelona Challenge from the European GNSS Agency (GSA), in 2015,
the Best Innovative Project of the Market Assessment Program (MAP) of EADA Business School, in 2016, the Award Isabel P. Trabal from Fundacióó Caixa d’Enginyers for its quality research during the Ph.D., in 2017, and the Best Ph.D. Thesis Award in remote sensing from the IEEE Geoscience and Remote Sensing (GRSS) Spanish Chapter, Spain, in 2019.
\end{IEEEbiography}

\begin{IEEEbiography}[{\includegraphics[width=1in,height=1.25in,clip]{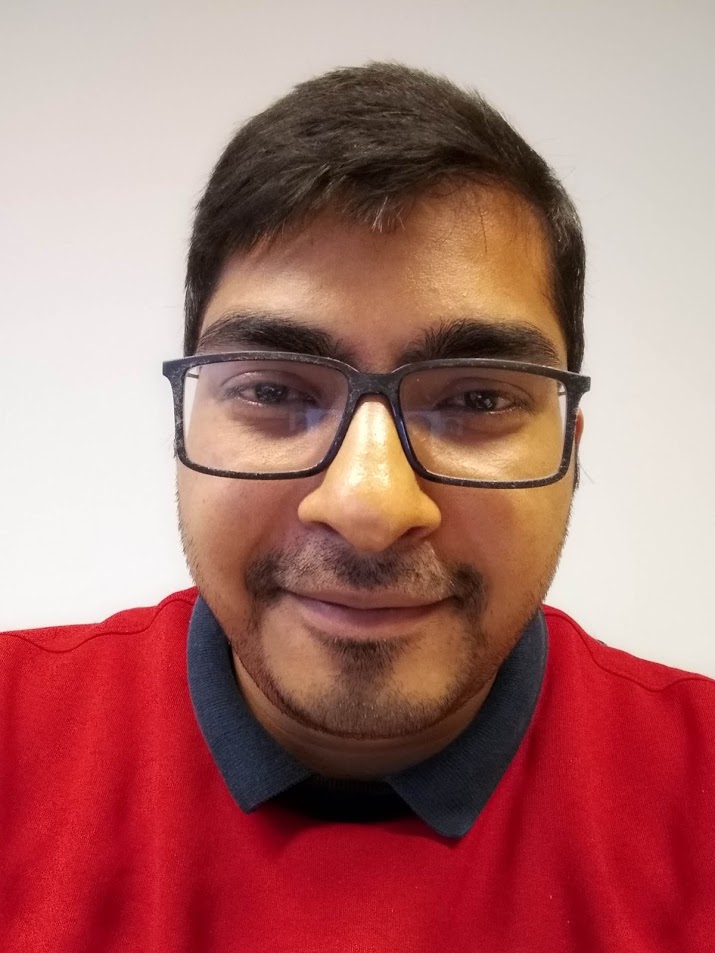}}]{\textbf{Sumit Kumar}} (S’14-M’19)  received the B.Tech and M.S. in Electronics $\&$ Communication Engineering from Gurukula Kangri University, Haridwar, India (2008) and the International Institute of Information Technology, Hyderabad, India (2014), respectively, and the Ph.D. from EURECOM (France) in 2019. Currently, he is working as a Research Associate at the Interdisciplinary Centre for Security, Reliability, and Trust (SnT),
University of Luxembourg. His research interests are in wireless communication, interference management, Integration of 5G with Non-Terrestrial-Networks and Software Defined Radio prototyping.

\end{IEEEbiography}

\begin{IEEEbiography}[{\includegraphics[width=1in,height=1.25in,clip]{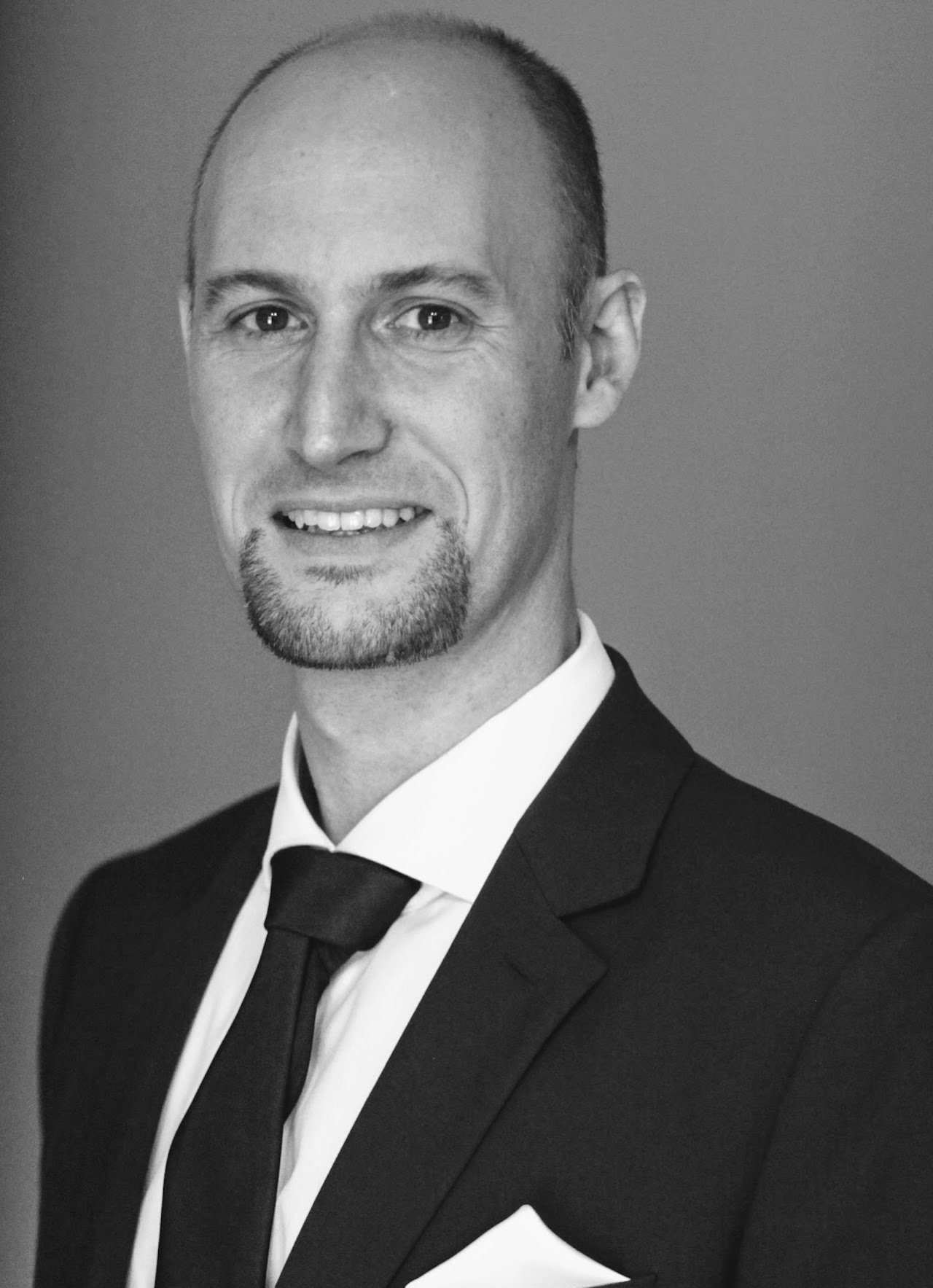}}]{\textbf{Symeon Chatzinotas}}(Fellow, IEEE) is currently Full Professor / Chief Scientist I and Head of the research group SIGCOM in the Interdisciplinary Centre for Security, Reliability and Trust, University of Luxembourg. In parallel, he is an Adjunct Professor in the Department of Electronic Systems, Norwegian University of Science and Technology and a Collaborating Scholar of the Institute of Informatics \& Telecommunications, National Center for Scientific Research “Demokritos”. 

In the past, he has lectured as Visiting Professor at the University of Parma, Italy and contributed in numerous R\&D projects for the Institute of Telematics and Informatics, Center of Research and Technology Hellas and Mobile Communications Research Group, Center of Communication Systems Research, University of Surrey.
He has received the M.Eng. in Telecommunications from Aristotle University of Thessaloniki, Greece and the M.Sc. and Ph.D. in Electronic Engineering from University of Surrey, UK in 2003, 2006 and 2009 respectively. 

He has authored more than 700 technical papers in refereed international journals, conferences and scientific books and has received numerous awards and recognitions, including the IEEE Fellowship and an IEEE Distinguished Contributions Award. He is currently in the editorial board of the IEEE Transactions on Communications, IEEE Open Journal of Vehicular Technology and the International Journal of Satellite Communications and Networking.
\end{IEEEbiography}

\clearpage
\onecolumn
\doublespacing
\fontsize{12}{14}\selectfont

\end{document}